\documentclass[aps,pra,onecolumn,showpacs]{revtex4}
\usepackage{amsmath,epsfig,amsfonts,graphicx,}
\usepackage{subfigure}

\begin{document}

\title{Spatial solitons under competing linear and nonlinear diffractions}
\author{Y. Shen, P. G. Kevrekidis, and N. Whitaker}
\affiliation{Department of Mathematics and Statistics, University of Massachusetts,
Amherst, MA 01003-4515, USA}
\author{Boris A. Malomed}
\affiliation{Department of Physical Electronics,\ School of Electrical Engineering,
Faculty of Engineering, Tel Aviv University, Tel Aviv 69978, Israel}

\begin{abstract}
We introduce a general model which augments the one-dimensional nonlinear
Schr\"{o}dinger (NLS) equation by nonlinear-diffraction terms competing with
the linear diffraction. The new terms contain two irreducible parameters and
admit a Hamiltonian representation in a form natural for optical media. The
equation serves as a model for spatial solitons near the supercollimation
point in nonlinear photonic crystals. In the framework of this model, a
detailed analysis of the fundamental solitary waves is reported, including
the variational approximation (VA), exact analytical results, and systematic
numerical computations. The Vakhitov-Kolokolov (VK) criterion is used to
precisely predict the stability border for the solitons, which is found in
an exact analytical form, along with the largest total power (norm) that the
waves may possess. Past a critical point, collapse effects are observed,
caused by suitable perturbations. Interactions between two identical
parallel solitary beams are explored by dint of direct numerical
simulations. It is
found that in-phase solitons merge into robust or collapsing pulsons,
depending on the strength of the nonlinear diffraction.
\end{abstract}

\maketitle


\section{Introduction}

It is well known that bright spatial solitons in optical media are supported
by the balance between the self-focusing nonlinearity and diffraction. A
great deal of attention has also been recently attracted to more special
cases, when the diffraction is subject to various forms of \textit{management%
}, so that the diffraction coefficient gradually vanishes or periodically
changes its sign~\cite{management}. In particular, it was shown in Ref.~\cite%
{Zhong} that related to such a setting with a vanishing diffraction
coefficient is another recently proposed class of models, in which stable
bright solitons are supported by the interplay of the ordinary diffraction
and self-defocusing nonlinearity, whose local strength grows at $%
r\rightarrow \infty $ at any rate faster than $r^{D}$, where $D$ is the
spatial dimension \cite{Barcelona}.

A physically important realization of the effectively weak diffraction,
which may therefore be sensitive to corrections that are usually negligible,
is the setting corresponding to the \textit{supercollimation} in photonic
crystals. This setting was demonstrated experimentally \cite{experiment} and
studied in detail theoretically \cite{theory}-\cite{xu}. Similar situations,
and the possibility of the existence of solitary waves in them, were also
analyzed in other photonic media, such as waveguiding arrays \cite{arrays},
cavities \cite{cavity}, and chains with the quadratic nonlinearity \cite%
{quadr-array}. Furthermore, it was proposed that similar schemes may be
implemented as well for acoustic waves in sonic crystals \cite{sonic-crystal}%
, and in Bose-Einstein condensates loaded into optical lattices \cite{BEC}.

The smallness of the diffraction coefficient near the supercollimation point
suggests that the corresponding model should be extended by adding nonlinear
corrections to the diffraction term in the usual nonlinear Schr\"{o}dinger
(NLS) equation. Such a generalization of the NLS equation was proposed in
Ref. \cite{xu}. In the framework of the generalized model, new properties of
spatial solitons were predicted---in particular, the existence of a maximum
total power, above which solitary beams do not exist, as well as inelastic
interactions between the beams. The objective of the present work is to
extend the model proposed in Ref. \cite{xu}, and study basic properties of
solitary waves in the extended system. The main modifications, in comparison
with the previously studied model, are that we propose the most general
nonlinear-diffraction terms which agree with the Hamiltonian structure of
the extended NLS equation (the term introduced in Ref. \cite{xu} cannot be
derived from a natural Hamiltonian), and that these general terms depend on
two irreducible parameters, rather than a single parameter, which was the
case in Ref. \cite{xu}.

The extended model is introduced in Section II, where we also present a
variational approximation (VA) \cite{VA} for the study of fundamental
solitary waves in this model, and some additional analytical results.
Numerical results for the shape and stability of the solitons, including the
comparison with predictions of the VA and interactions between solitons
forming out-of-phase and in-phase pairs, are reported in Section III. The
paper is concluded by Section IV.

\section{The model and analytical framework}

Our aim is to derive a general NLS model including different contributions
to the nonlinear diffraction. This is done by developing an extended version
of the model proposed in Ref. \cite{xu} for the description of the
supercollimation in photonic crystals. 
We aim to extend the model by using a Lagrangian formulation that produces
terms similar to the one introduced in Ref. \cite{xu}, along with additional
terms invoked by the underlying Lagrangian/Hamiltonian structure.

Our starting point is the Lagrangian density for an extended NLS equation
including the nonlinear diffraction in a general form, $\mathcal{L}=\left(
i/2\right) (q^{\ast }q_{\xi }-qq_{\xi }^{\ast })-\mathcal{H}$, where $q$ is
the local wave amplitude, the asterisk stands for the complex conjugation, $%
\xi $ is the propagation distance, $\eta $ is the transverse coordinate in
the planar waveguide, and the Hamiltonian density is
\begin{equation}
\mathcal{H}=|q_{\eta }|^{2}-\frac{1}{2}|q|^{4}-\frac{\beta }{2}%
|q|^{2}|q_{\eta }|^{2}-\frac{\gamma }{4}\left[ \left( {q^{\ast }}\right)
^{2}q_{\eta }^{2}+q^{2}\left( {q_{\eta }^{\ast }}\right) ^{2}\right] ,
\label{H}
\end{equation}%
with real coefficients $\beta $ and $\gamma $. Varying the Lagrangian, $%
L=\int_{-\infty }^{+\infty }\mathcal{L}d\eta $, gives rise to the
generalized NLS equation,%
\begin{equation}
iq_{\xi }+q_{\eta \eta }+|q|^{2}q-\frac{\beta }{2}\left( |q|^{2}q_{\eta \eta
}+q^{\ast }q_{\eta }^{2}\right) +\frac{\gamma }{2}\left( q^{\ast }q_{\eta
}^{2}-2q|q_{\eta }|^{2}-q^{2}q_{\eta \eta }^{\ast }\right) =0.  \label{ele}
\end{equation}%
This constitutes an extension of Eq.~(3) from Ref. \cite{xu}. In particular,
the term $-\left( \beta /2\right) q^{\ast }q_{\eta }^{2}$, which did not
appear in Ref. \cite{xu}, emerges along with its counterpart, $-\left( \beta
/2\right) |q|^{2}q_{\eta \eta }$, that accounted for the nonlinear
diffraction in Ref. \cite{xu}, in the framework of the self-consistent
Lagrangian/Hamiltonian formulation. Additionally, the terms proportional to $%
\gamma $, which were absent in the equation adopted in Ref. \cite{xu},
represent contributions to the nonlinear diffraction at the same order as
the terms $\sim \beta $, hence they should be included too. The remaining
terms in Eq. (\ref{ele}) correspond to the ordinary NLS equation.
Dynamical invariants of Eq. (\ref{ele}) are the Hamiltonian, $%
H=\int_{-\infty }^{+\infty }\mathcal{H}d\eta $, momentum $M=i\int_{-\infty
}^{+\infty }q_{\eta }^{\ast }qd\eta $, and the total power (norm), which we
define as%
\begin{equation}
P\equiv \pi ^{-1/2}\int_{-\infty }^{+\infty }|q(\eta )|^{2}d\eta .  \label{p}
\end{equation}

It is relevant to mention another equation including the nonlinear
diffraction, which was derived as a continuum limit of the discrete Salerno
model \cite{Zaragoza} (see also Ref. \cite{additional}):%
\begin{equation}
iq_{\xi }+\left( 1-m|q|^{2}\right) q_{\eta \eta }+2\left( 1-m\right)
|q|^{2}q=0,  \label{Zara}
\end{equation}%
with parameter $m$ taking values $0\leq m<1$. In fact, it is the same
equation as the one proposed in Ref. \cite{xu}, with $\beta \equiv m/\left(
1-m\right) $. As shown in Ref. \cite{Zaragoza}, Eq. (\ref{Zara}) conserves
the norm and Hamiltonian,%
\begin{eqnarray}
P_{m} &=&\int_{-\infty }^{+\infty }\left[ \ln \left( 1-m|q|^{2}\right) %
\right] d\eta , \\
H_{m} &=&\int_{-\infty }^{+\infty }\left[ \left\vert q_{\eta }\right\vert
^{2}+2\left( \frac{1}{m}-1\right) |q|^{2}+\frac{2}{m^{2}}\ln \left(
1-m|q|^{2}\right) \right] d\eta .
\end{eqnarray}%
However, for the purpose of modeling optical media, these norm and
Hamiltonian do not seem natural, unlike those given by Eqs. (\ref{p}) and (%
\ref{H}), therefore Eq. (\ref{ele}) provides, arguably, a more appropriate
model than Eq. (\ref{Zara}).

Following the lines of Ref. \cite{xu}, we do not include the fourth-order
linear diffraction, that would be represented by a term $\sim q_{\eta \eta
\eta \eta }$ in Eq. (\ref{ele}), assuming that the equation applies to
relatively broad beams, and our objective is to focus on effects of
nonlinear diffraction. Further, as in Refs. \cite{Zaragoza} and \cite{xu},
we focus here on the case of $\beta >0$, which implies that the nonlinear
diffraction competes with its linear counterpart, as in the opposite case a
nonlinear enhancement of the diffraction does not lead to particularly
noteworthy effects.

A straightforward virial estimate demonstrates that Eq. (\ref{ele}) with $%
\beta >0$ gives rise to a supercritical collapse (catastrophic
self-compression of the wave field) \cite{Berge}; the collapse will be
arrested if the fourth-order linear diffraction is included too. In other
words, it is a \textit{weak collapse}, so called because only a small part
of the total power (norm) of the field is involved into the
self-compression, the rest being scattered away (the collapse is called
\textit{strong}, involving an essential part of the total power, in the
\textit{critical case}, which is driven, for instance, by the quintic
self-focusing term in the one-dimensional NLS equation \cite{Berge,Kuz}).

Stationary solutions to Eq.~(\ref{ele}) are sought in the usual form, $q(\xi
,\eta )=e^{ik\xi }Q(\eta )$, where $Q(\eta )$ is a real function satisfying
the ordinary differential equation
\begin{equation}
-kQ+\frac{d^{2}Q}{d\eta ^{2}}-\frac{1}{2}\left( \beta +\gamma \right) \left[
Q^{2}\frac{d^{2}Q}{d\eta ^{2}}+Q\left( \frac{dQ}{d\eta }\right) ^{2}\right]
+Q^{3}=0.  \label{Q}
\end{equation}%
Rescaling the field amplitude and coordinate as%
\begin{equation}
Q\equiv \sqrt{k}\tilde{Q},~x\equiv \sqrt{k}\eta  \label{Qx}
\end{equation}%
casts Eq. (\ref{Q}) into a normalized form,
\begin{equation}
-\tilde{Q}+\frac{d^{2}\tilde{Q}}{dx^{2}}-\frac{1}{2}(\beta +\gamma )k\left[
\left( \frac{d\tilde{Q}}{dx}\right) ^{2}\tilde{Q}+{\tilde{Q}}^{2}\frac{d^{2}%
\tilde{Q}}{dx^{2}}\right] +\tilde{Q}^{3}=0,  \label{Qtilde}
\end{equation}%
which is more appropriate for the analysis, as it contains a single free
parameter, $(\beta +\gamma )k$. Accordingly, the $P(k)$ dependence,
following from Eqs. (\ref{p}) and (\ref{Qx}), takes the form of%
\begin{equation}
P=\sqrt{k}\tilde{P}\left( \left( \beta +\gamma \right) k\right) ,
\label{tilde}
\end{equation}%
where $\tilde{P}\equiv \pi ^{-1/2}\int_{-\infty }^{+\infty }\left( \tilde{Q}%
(x)\right) ^{2}dx$ is considered as a function of $\left( \beta +\gamma
\right) k$. Further, it follows from Eq. (\ref{tilde}) that the \textit{%
cutoff} value of the propagation constant, $k_{\mathrm{co}}$, which is a
border of the soliton stability region according to the VK criterion, $%
dP/dk=0$, is determined by the equation $\tilde{P}+2\left( \beta +\gamma \right)
k_{\mathrm{co}}\tilde{P}^{\prime }\left( \left( \beta +\gamma \right) k_{%
\mathrm{co}}\right) =0$ (the prime stands here for the derivative). It
follows from here that the cutoff value of the power, $P_{\max }$, above
which the solitons do not exist (see the top right panel in Fig. \ref{WA}
below), together with the corresponding soliton's peak power and width (that
represent, respectively, the largest peak power and smallest width that
stable solitons may have), depend on the nonlinear-diffraction parameters as
follows:
\begin{eqnarray}
k_{\mathrm{co}} &=&C_{k}\left( \beta +\gamma \right) ^{-1},~P_{\max
}=C_{P}\left( \beta +\gamma \right) ^{-1/2},~  \notag \\
A_{\max }^{2} &=&C_{A}\left( \beta +\gamma \right) ^{-1},~W_{\min
}=C_{W}\left( \beta +\gamma \right) ^{1/2},  \label{co}
\end{eqnarray}%
where numerical found constants are $C_{k}\approx 0.66$, $C_{P}\approx 2.35$%
, $C_{A}\approx 1.28$, and $C_{W}\approx 2.42$. For comparison, it is
relevant to mention that in the model based on Eq. (\ref{Zara}) the cutoff
value of the propagation constant,
\begin{equation}
k_{\mathrm{co}}=\left( 1-m\right) /m\equiv \beta ^{-1},  \label{Aragon}
\end{equation}%
bounds the \emph{existence}, rather than stability, of the solitons \cite%
{Zaragoza}. Note that Eq. (\ref{Aragon}) reveals the same scaling of $k_{%
\mathrm{co}}$ vs. $\beta $ as Eq. (\ref{co}) demonstrates for $k_{\mathrm{co}%
}$ vs. $\left( \beta +\gamma \right) $.


Numerical solutions obtained with the help of Eq. (\ref{Qtilde}) are
produced in the next section.
Before turning to that, we present the VA for soliton solutions. This
approach is well known to produce relevant predictions for many dynamical
effects exhibited by solitons \cite{VA}, such the influence of phase
modulations on the formation of solitons \cite{Andrei}. To this end, we
adopt the Gaussian ansatz with amplitude $A$ and width $W$,
\begin{equation}
Q(\eta )=A\exp \left[ -\eta ^{2}/\left( 2W^{2}\right) \right] ,
\label{Gauss}
\end{equation}%
which is relevant for small values of the propagation constant $k$ [see Fig. (%
\ref{Qfig}) below]. With the increase of $k$, the numerically found solution
develops a peakon-like shape, hence we expect the VA to fail at larger $k$.

Inserting the ansatz (\ref{Gauss}) into the Lagrangian density corresponding
to Eq. (\ref{Q}),
\begin{equation}
\mathcal{L}_{Q}=-kQ^{2}-\left( \frac{d{Q}}{d\eta }\right) ^{2}+\frac{\beta
-\gamma }{2}Q^{2}\left( \frac{d{Q}}{d\eta }\right) ^{2}+\frac{1}{2}Q^{4},
\end{equation}%
we calculate the effective Lagrangian,
\begin{equation}
L_{\mathrm{eff}}\equiv \int_{-\infty }^{\infty }\mathcal{L}_{Q}d\eta =-\sqrt{%
\pi }kA^{2}W-\frac{\sqrt{\pi a}}{2}\frac{A^{2}}{W}+\left( \beta -\gamma
\right) A^{4}\frac{\sqrt{2\pi }}{16W}+\frac{1}{2}\sqrt{\frac{\pi }{2}}A^{4}W,
\label{Leff}
\end{equation}%
where $A^{2}$ may be replaced by the total power (\ref{p}) corresponding to
ansatz (\ref{Gauss}), which is $P=A^{2}W$. This substitution yields
\begin{equation}
\frac{4}{\sqrt{\pi }}L_{\mathrm{eff}}=-4kP-\frac{2P}{W^{2}}+\frac{\beta
-\gamma }{2\sqrt{2}}\frac{P^{2}}{W^{3}}+\sqrt{2}\frac{P^{2}}{W}.  \label{eff}
\end{equation}%
Then, the Euler-Lagrange equations associated with stationary solutions
within the framework of the VA are given by
\begin{eqnarray}
\frac{\partial {L_{\mathrm{eff}}}}{\partial {P}}=0 &\Rightarrow &-4k-\frac{2%
}{W^{2}}+\frac{\beta -\gamma }{\sqrt{2}}\frac{P}{W^{3}}+2\sqrt{2}\frac{P}{W}%
=0, \\
\frac{\partial {L}_{\mathrm{eff}}}{\partial {W}}=0 &\Rightarrow &\frac{1}{W}-%
\frac{3(\beta -\gamma )}{8\sqrt{2}}\frac{P}{W^{2}}-\frac{1}{2\sqrt{2}}P=0.
\end{eqnarray}%
Eliminating $P$ here, we arrive at an equation for the width,
\begin{equation}
8kW^{4}+6[k(\beta -\gamma )-2]W^{2}-(\beta -\gamma )=0.  \label{bi}
\end{equation}%
Obviously, this equation can be solved analytically. For instance, the
physical solution for $\beta =\gamma $ is $W^{2}=3/(2k)$.

In what follows below, the linear stability of the solitary-wave solutions
is analyzed by means of numerical methods, with the aim to identify
parameter regions where the solutions are dynamically stable or unstable. To
this end, we consider perturbations of the stationary solutions in the form
of $q(\xi ,\eta )=\left[ Q(\eta )+\varepsilon \left( v(\eta )+iw(\eta
)\right) \right] e^{ik\xi }$, with an infinitesimally small perturbation
amplitude $\varepsilon $. Upon the substitution of this into Eq. (\ref{ele})
and linearization, we obtain the following equations:
\begin{eqnarray}
v_{\xi }=kw-w_{\eta \eta } &-&\beta \left( -Q^{2}w_{\eta \eta }+Q_{\eta
}^{2}w-2QQ_{\eta }w_{\eta }\right)  \notag \\
&-&\gamma \left( 2QQ_{\eta }w_{\eta }-3Q_{\eta }^{2}w+Q^{2}w_{\eta \eta
}-2QQ_{\eta \eta }w\right) -Q^{2}w,  \notag \\
w_{\xi }=-kv +v_{\eta \eta } & + &\beta \left( -Q^{2}v_{\eta \eta
}-2Q_{\eta \eta }Qv-Q_{\eta }^{2}v-2QQ_{\eta }v_{\eta }\right)  \notag \\
&+&\gamma \left( -2QQ_{\eta }v_{\eta }-Q_{\eta }^{2}v-Q^{2}v_{\eta \eta
}-2QQ_{\eta \eta }v\right) +3Q^{2}v,  \label{stability}
\end{eqnarray}%
which can be written in a symbolic form as $v_{\xi }=-\hat{L}_{1}w,\ w_{\xi
}=\hat{L}_{2}v$, with accordingly defined operators $\hat{L}_{1,2}$. Looking
for the stability eigenvalues $\lambda ,$ so that $v(\xi ,\eta )=e^{\lambda
\xi }\tilde{v}(\eta ),\ w(\xi ,\eta )=e^{\lambda \xi }\tilde{w}(\eta )$, we
arrive at the linear-stability problem in the form of $\lambda ^{2}\tilde{v}%
=-L_{1}L_{2}\tilde{v}$, and $\lambda ^{2}\tilde{w}=-L_{2}L_{1}\tilde{w}$.
The solution of the linear stability problem is reported in the following
section.

A well-known necessary stability condition relevant to fundamental solitary
waves in models with the self-focusing nonlinearity is given by the VK
criterion \cite{VK}-\cite{Kuz}, $dP/dk>0$. 
It is shown below that this criterion applies to the
present model, being also sufficient for the stability of the solitons.

\section{Numerical results}

\begin{figure}[tbp]
\centering
\includegraphics[width=.6\textwidth]{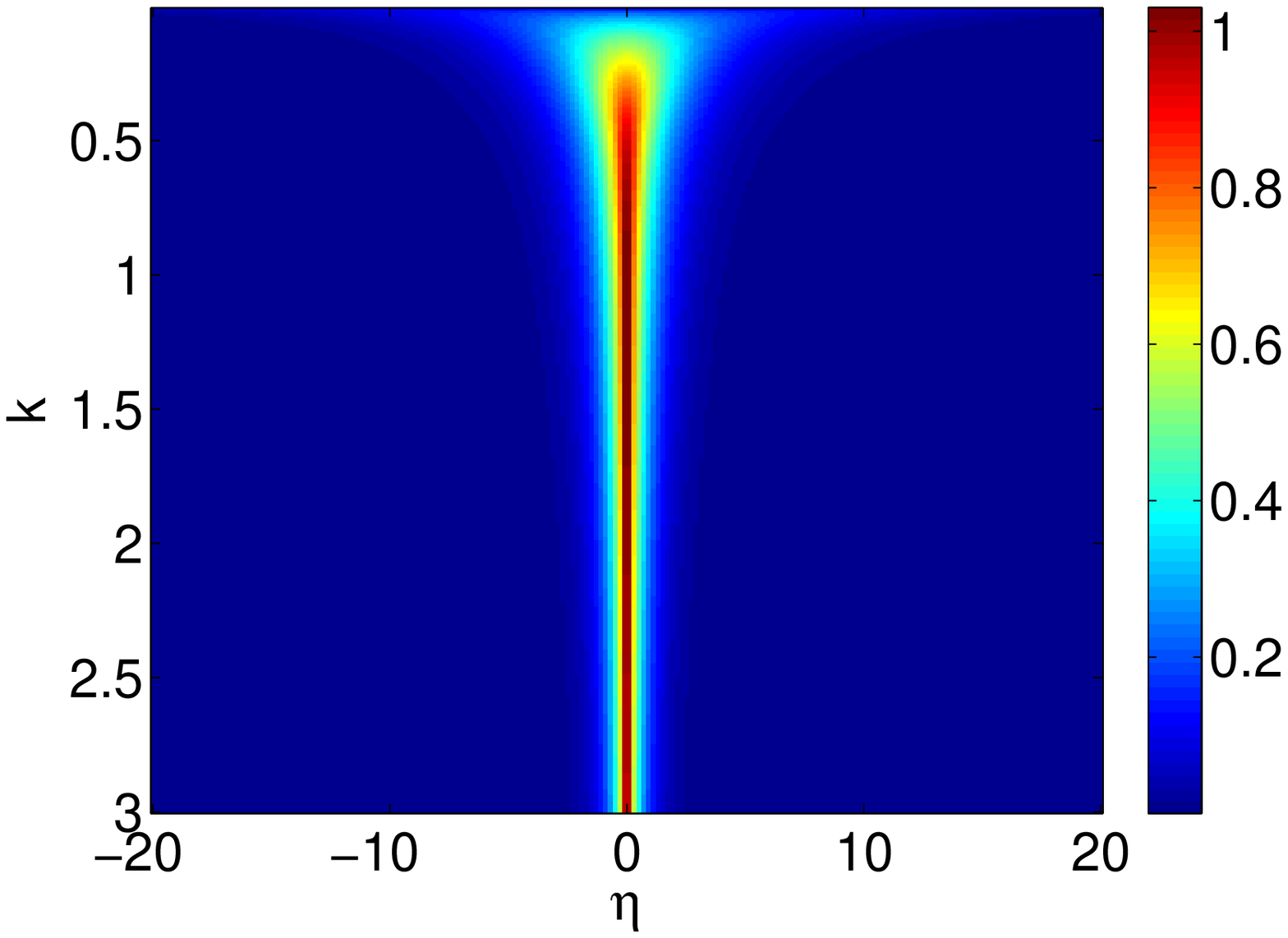} \includegraphics[width=.4%
\textwidth]{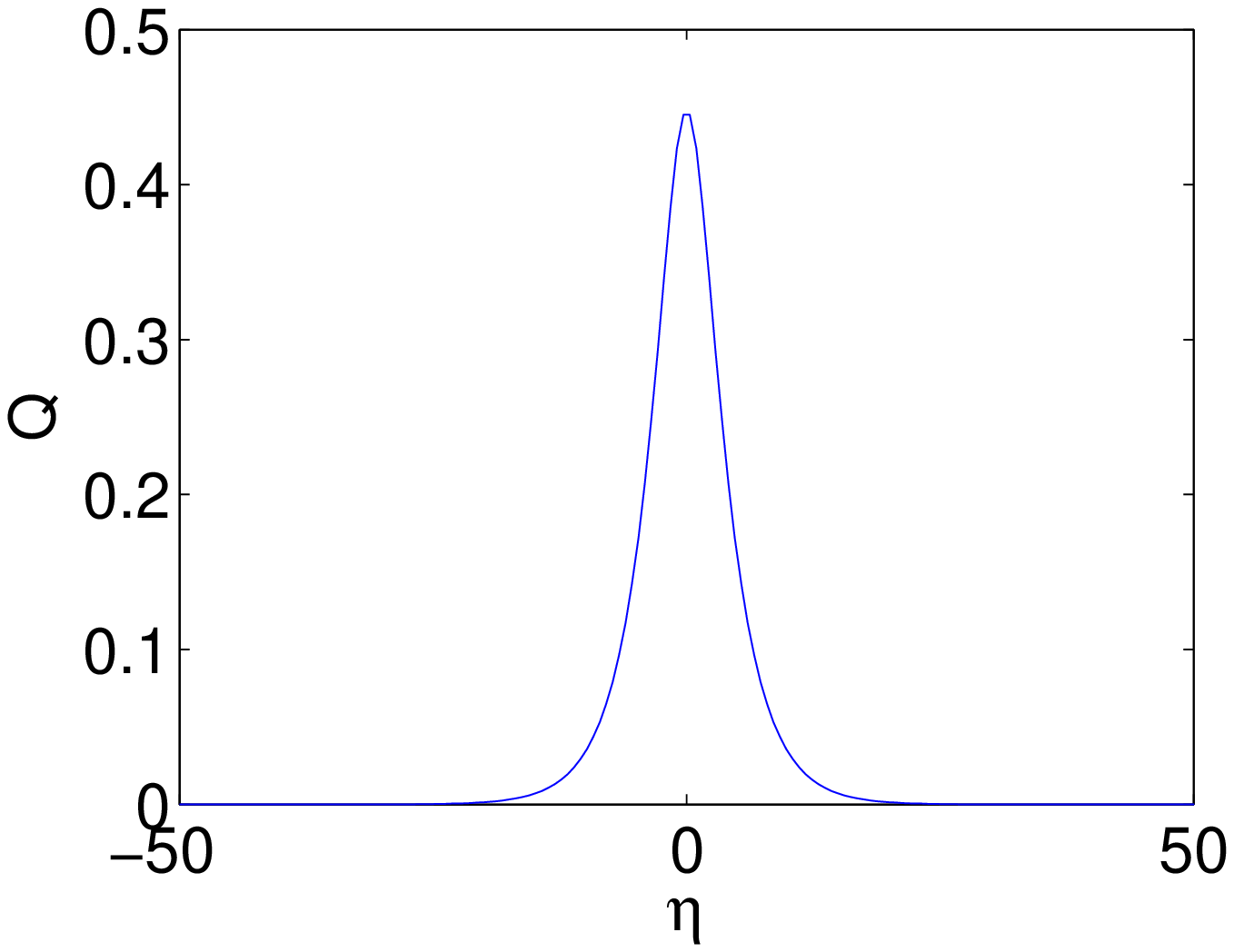} \includegraphics[width=.4\textwidth]{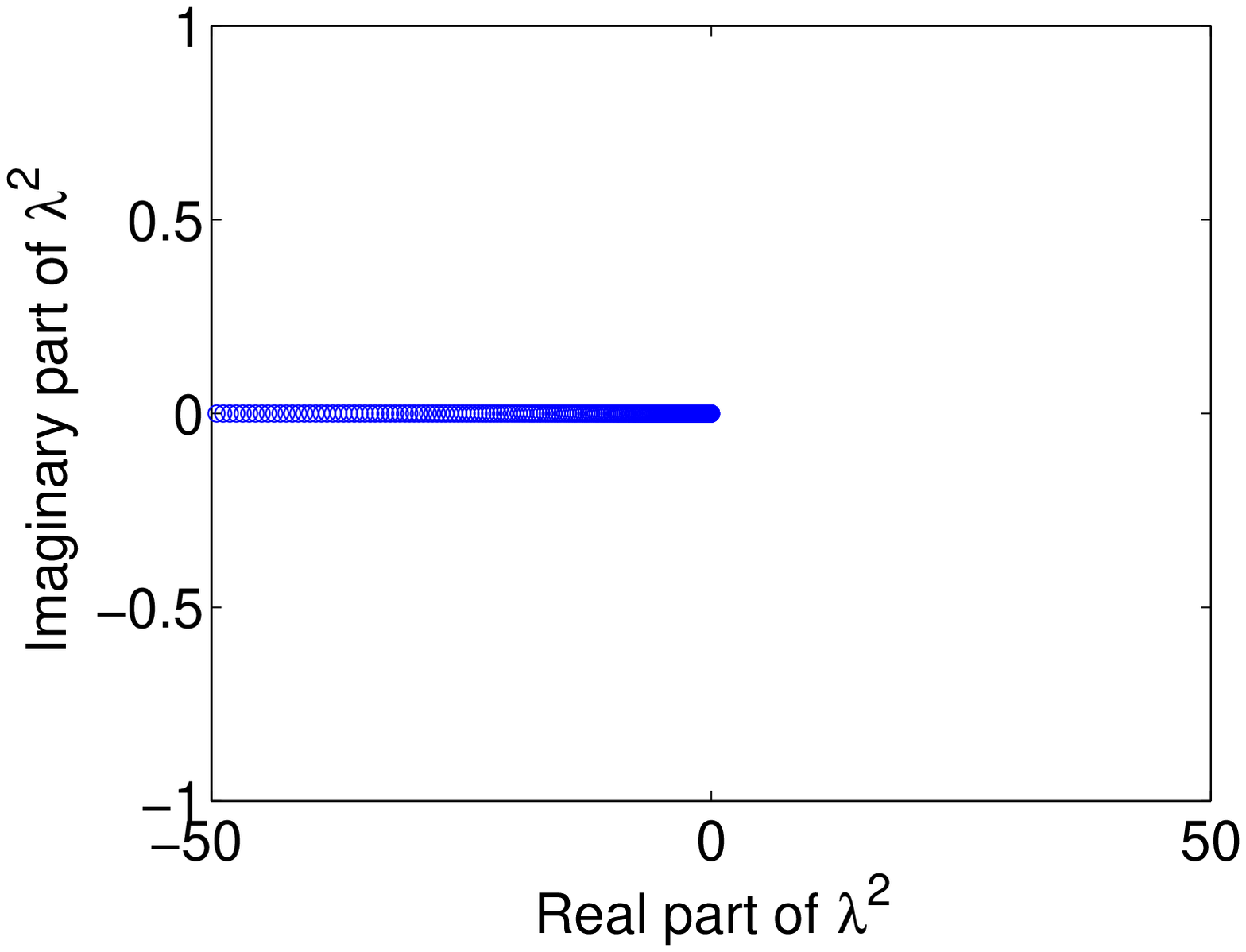} %
\includegraphics[width=.4\textwidth]{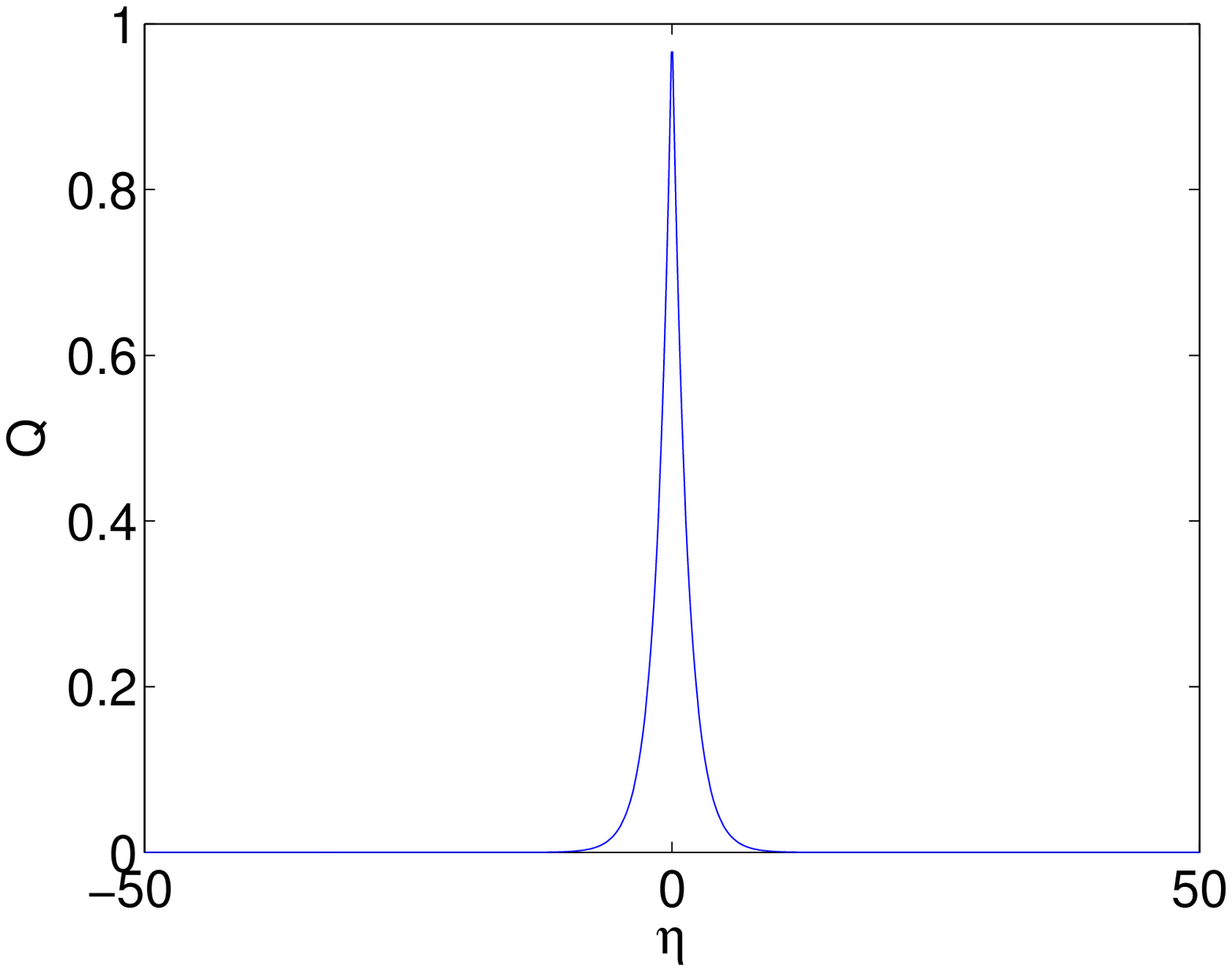} \includegraphics[width=.4%
\textwidth]{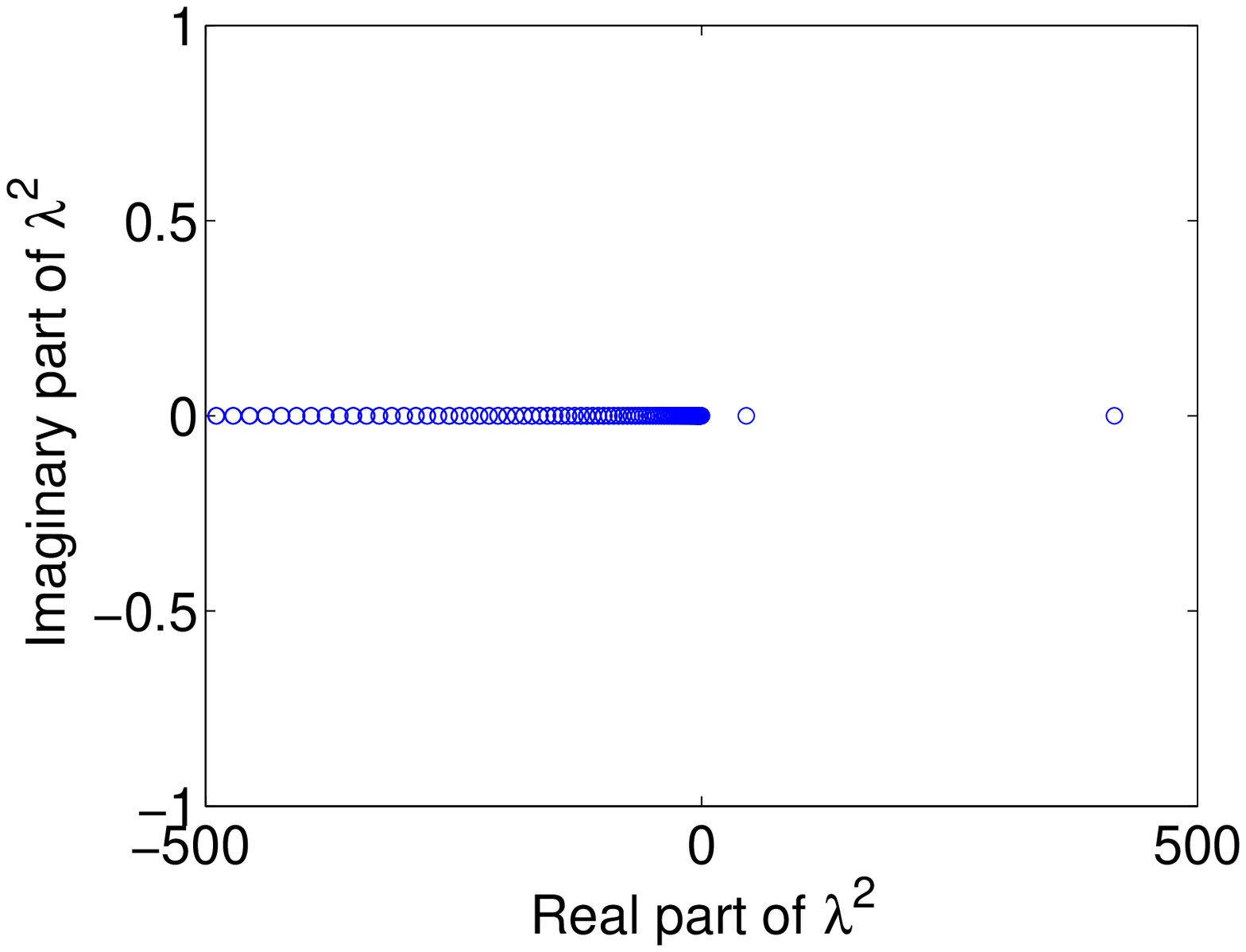}
\caption{(Color online) The top panel shows a sequence of solution profiles (%
$Q$) obtained with the increase of $k$ (recall $\protect\eta $ is the
spatial coordinate). The middle and bottom panels display the profile of $Q$%
, and the corresponding eigenvalues of the linear stability problem, for $%
k=0.1$ and $k=0.5$, respectively (accordingly, the former soliton is stable,
while the latter one is not). In this figure and below, the results are
displayed for $\protect\beta =1,\protect\gamma =1$, unless otherwise noted.}
\label{Qfig}
\end{figure}

We now turn to numerical solutions, and their comparison with results of the
VA. Fig. \ref{Qfig} presents the numerically obtained solutions of Eq. (\ref%
{Qtilde}) as a function of the propagation constant, $k$, for $\beta =1\ $%
and $\gamma =1$ in Eqs. (\ref{Q}) and (\ref{Qtilde}). To obtain these
solutions, the Newton's method was used for solving the respective
boundary-value problem, with vanishing (i.e., homogeneous
Dirichlet) boundary conditions and the
derivatives calculated by means of the centered-finite-difference
approximation. Our initial guess for small $k$, i.e., small values of the
coefficient in front of the nonlinear-diffraction terms in Eq. (\ref{Qtilde}%
), was $\tilde{Q}(x)=\sqrt{2}\mathrm{sech}~x$, i.e., the stationary solution
of the NLS equation, and we used the parametric continuation in $k$ (for
each subsequent step in $k$, the previous solution was used as the initial
guess). Naturally, it is observed that, as $k$ increases, the nonlinearity
plays a more significant role and the relevant solution becomes narrower
(i.e., with a smaller width) and taller (with a larger amplitude).

\begin{figure}[tbp]
\centering
\includegraphics[width=.4\textwidth]{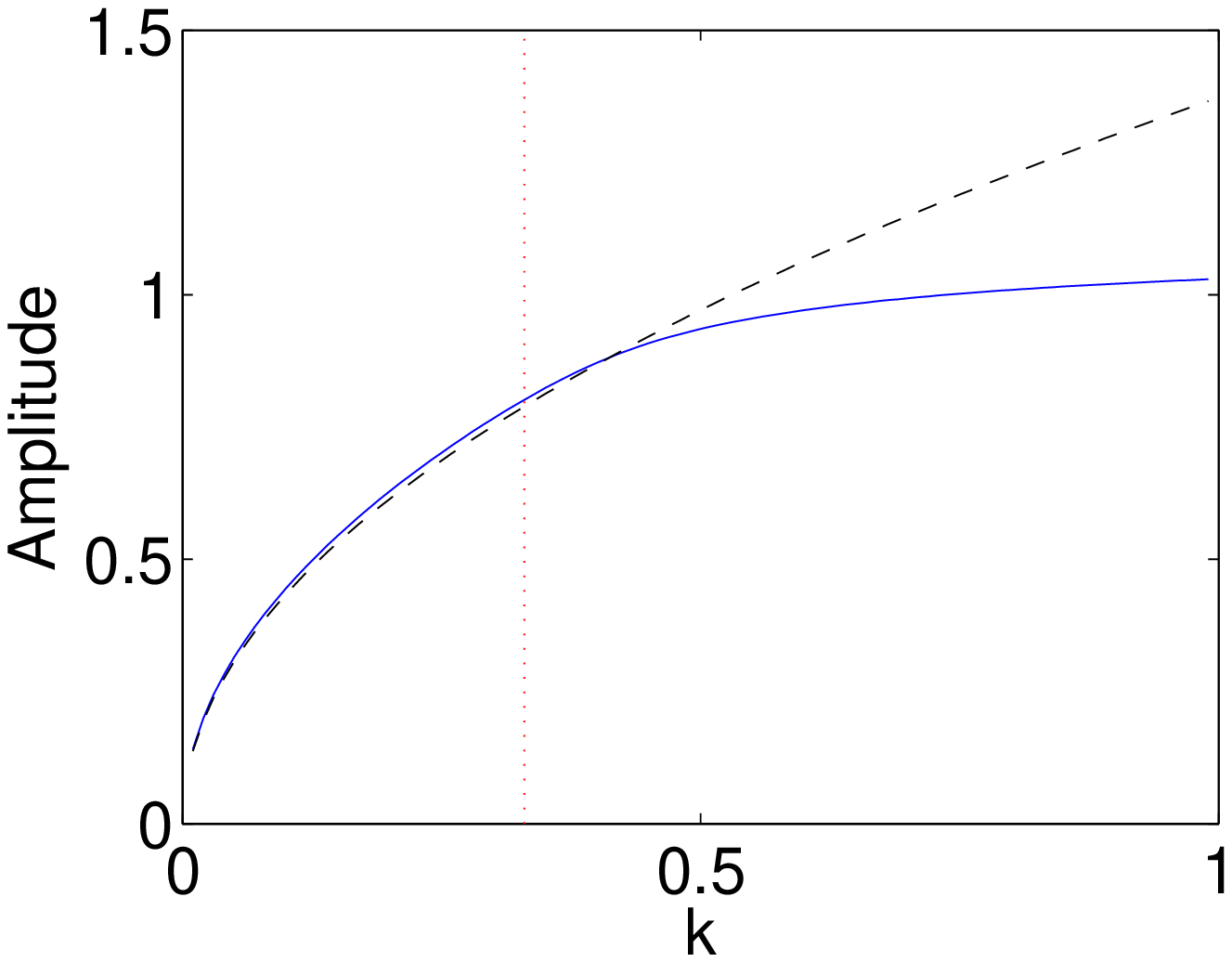} %
\includegraphics[width=.4\textwidth]{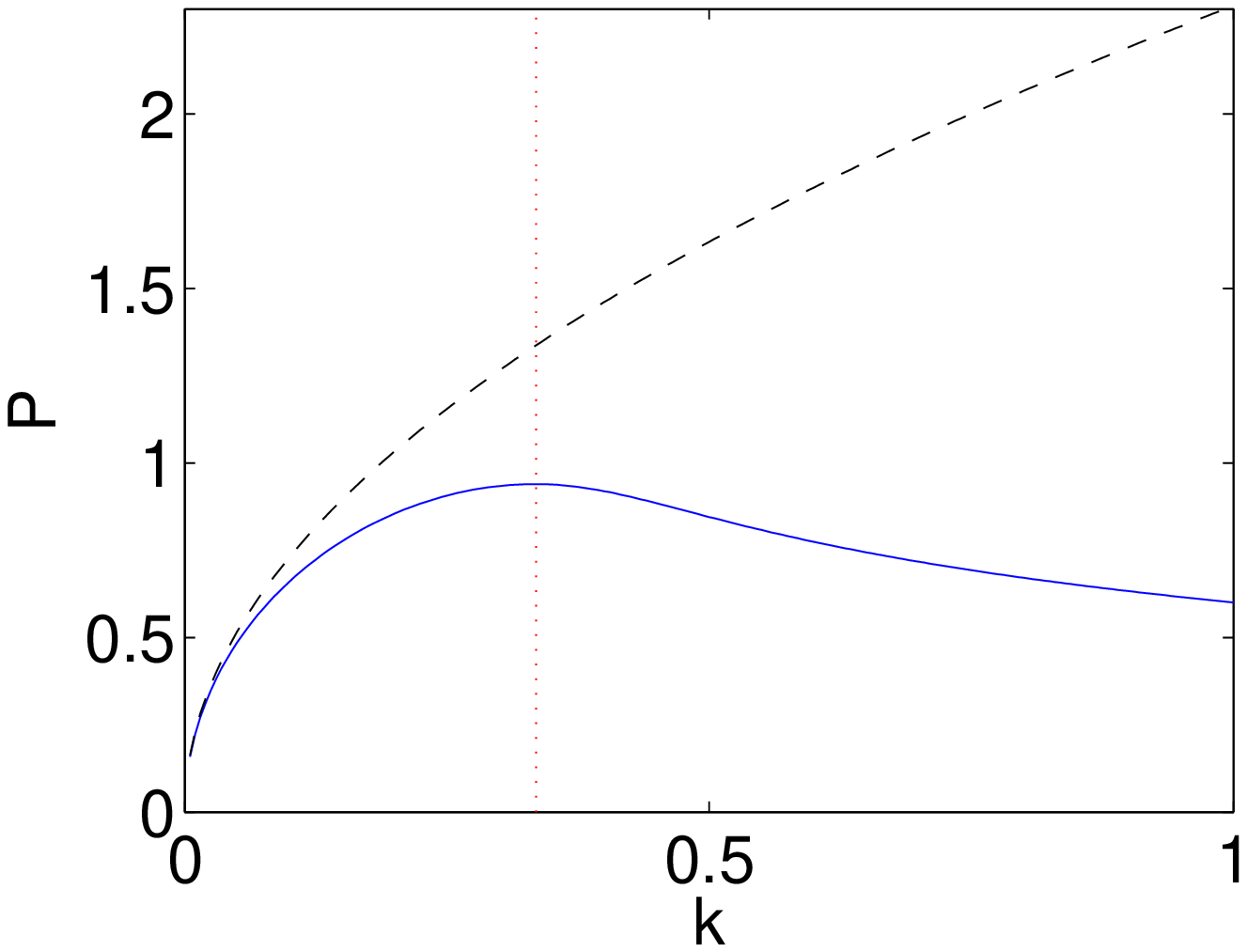} \includegraphics[width=.4%
\textwidth]{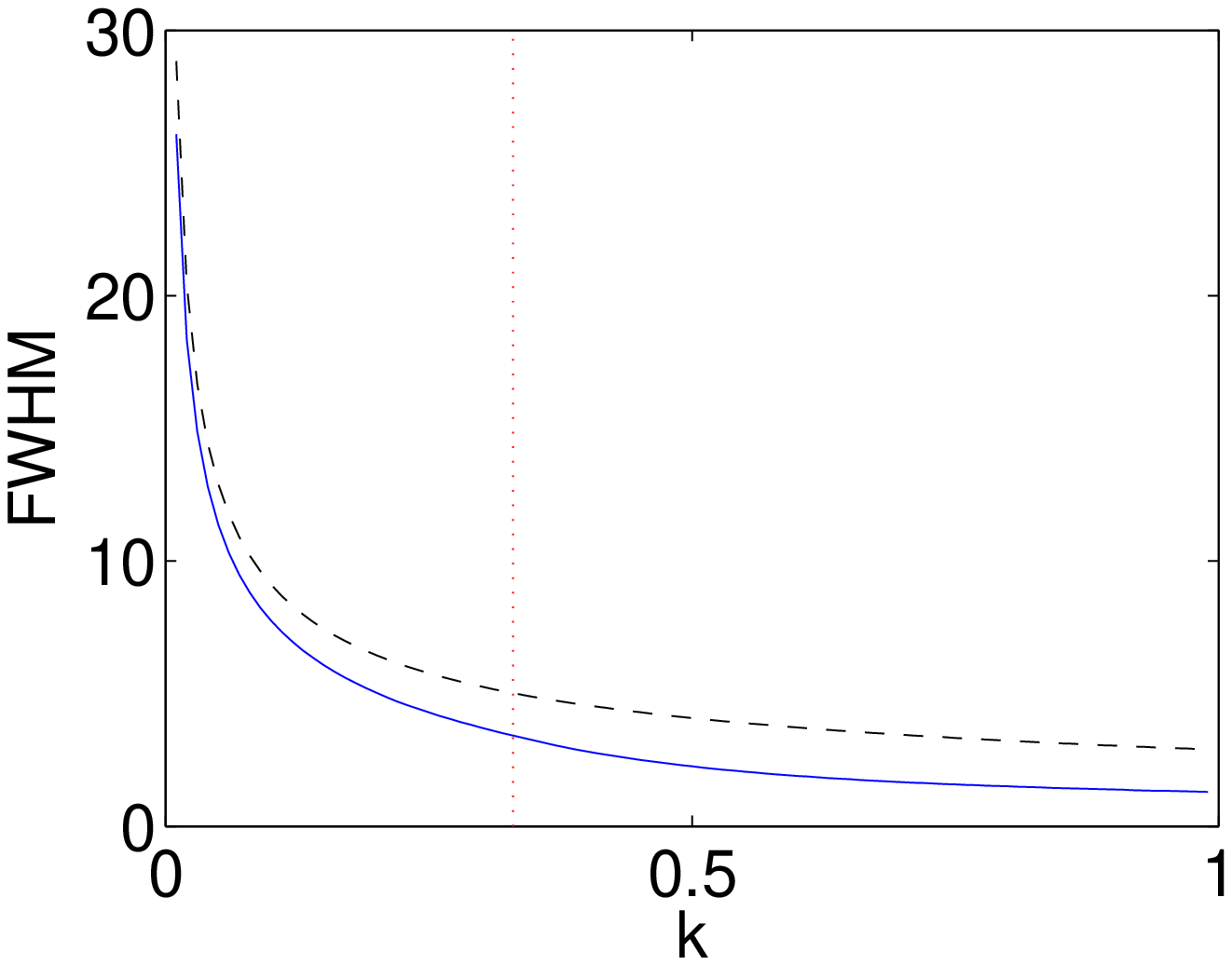} \includegraphics[width=.4\textwidth]{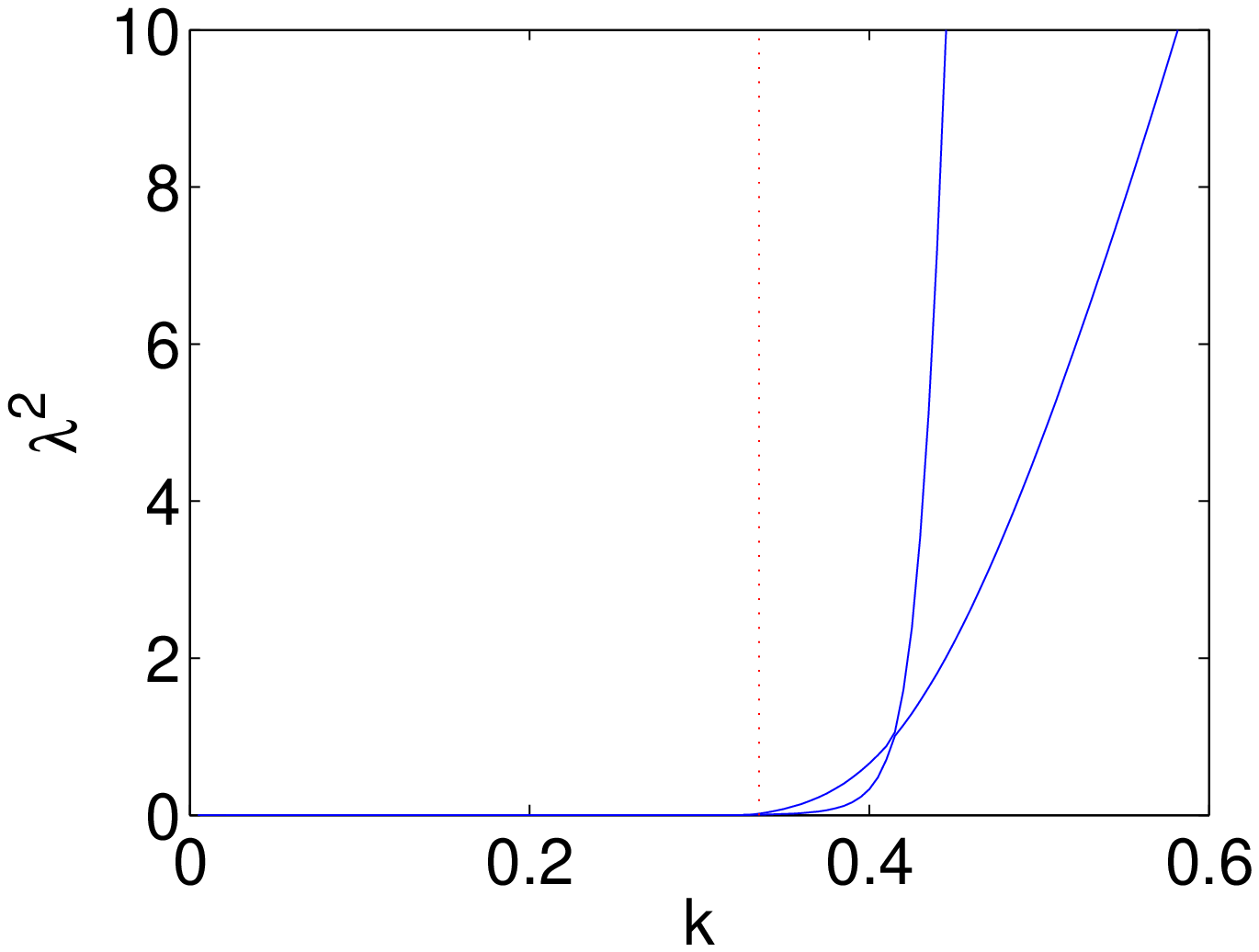}
\caption{(Color online) Characteristics of the soliton family with $\protect%
\beta =1,\ \protect\gamma =1$: The left top panel shows the amplitude vs. $k$%
, the right top is the power vs. $k$, the bottom left is the FWHM vs. $k$,
and the bottom right panel displays the two squared eigenvalues that
determine the instability of the solitary waves. The dashed and solid (blue)
lines represent, respectively, the predictions of the variational
approximation and numerical results. The vertical dotted (red) line marks $%
k=0.33$, at which the numerically found total power reaches its maximum, the
solitons being unstable at $k>0.33$, in the precise agreement with the VK
criterion and with Eq. (\protect\ref{co}). }
\label{WA}
\end{figure}

Figure \ref{WA} presents basic properties of the numerically obtained
solutions, and the comparison with their counterparts produced by the VA.
Particular characteristics of the soliton solutions that are displayed here
are the amplitude, total power, and full width at half maximum (FWHM) [the
total power is defined according to Eq. (\ref{p}), cf. Eq. (\ref{tilde})].
It is observed that, as expected, the VA is accurate for small values of $k$%
; however, it does not provide an adequate description for sharply-peaked
solutions at larger $k$, and it does not predict either the existence border
for the solitons, that corresponds to the the largest possible value of the
total power, $P=P_{\max }$, nor the stability-cutoff value of the
propagation constant, $k=k_{\mathrm{co}}$, which are evident in the top
right panel of Fig. \ref{WA}. However, as shown above, these values can be
found in the exact analytical form given by Eq. (\ref{co}), which is
unrelated to the VA. In fact, the wave packets suffer collapse at $%
P>P_{\max }$. It is relevant to mention that curves $A(k)$ and $P(k)$, which
are displayed in the left and right top panels of Fig. \ref{WA}, are
qualitatively similar to the respective dependences reported in the inset to
Fig. 2(d) and Fig. 2(b) of Ref. \cite{xu}, for the equation which is
tantamount to Eq. (\ref{Zara}).

Two squared eigenvalues which, for sufficiently large $k$, give rise to
instabilities are also shown in Fig. \ref{WA}, as a function of $k$. It
should be kept in mind that there is always a vanishing eigenvalue in the
system due to the phase (gauge) invariance. At $k>k_{\mathrm{co}}=0.33$,
when the monotonicity of the power of the solution changes, two eigenvalue
pairs turn positive (i.e., $\lambda ^{2}>0$), leading to instability, in
accordance with the VK criterion and with analytical results (\ref{co}).
Thus, the sharp-peaked solutions found at large $k$ are unstable.

\begin{figure}[tbp]
\centering
\includegraphics[width=.45\textwidth]{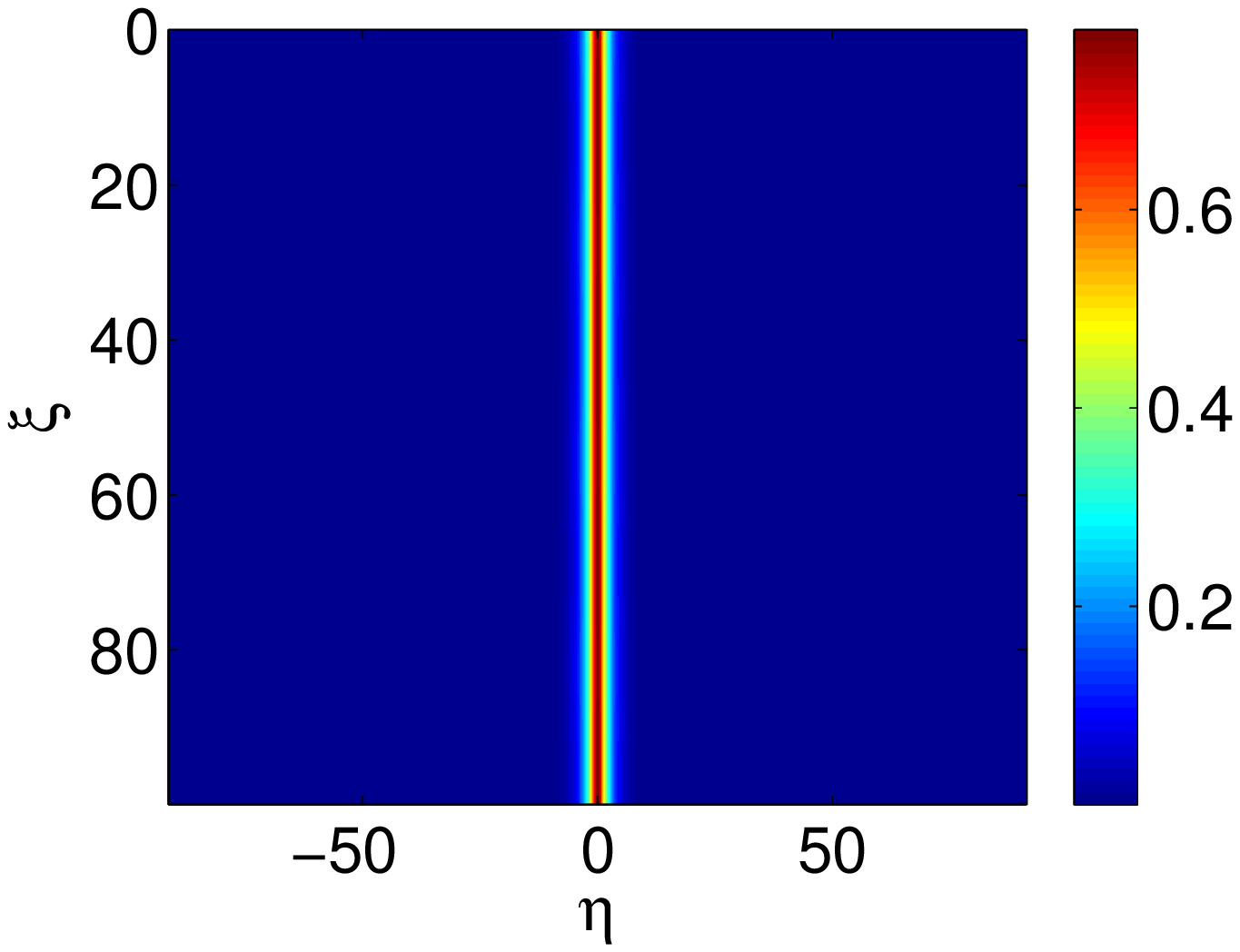} \includegraphics[width=.45%
\textwidth]{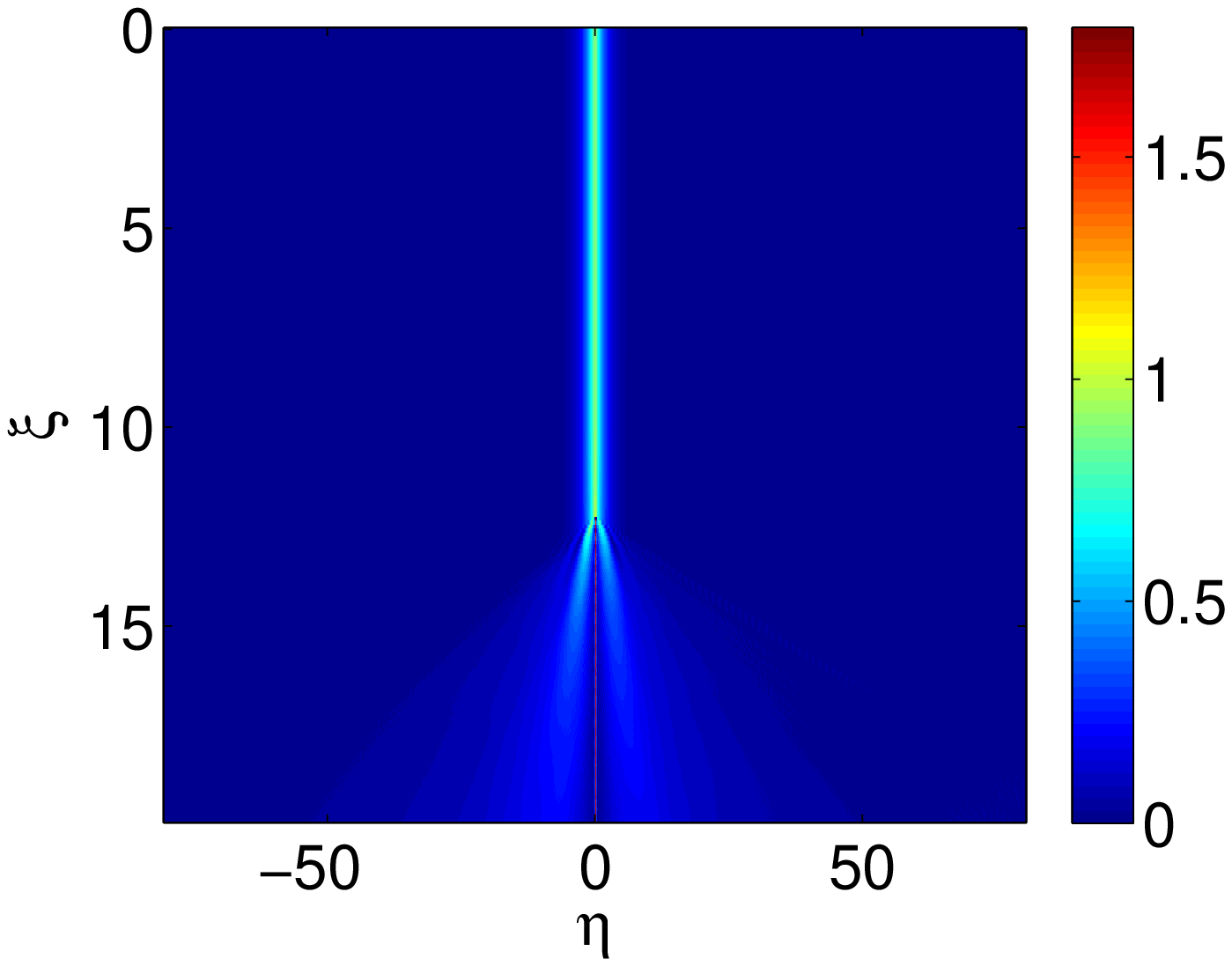} \includegraphics[width=.45%
\textwidth]{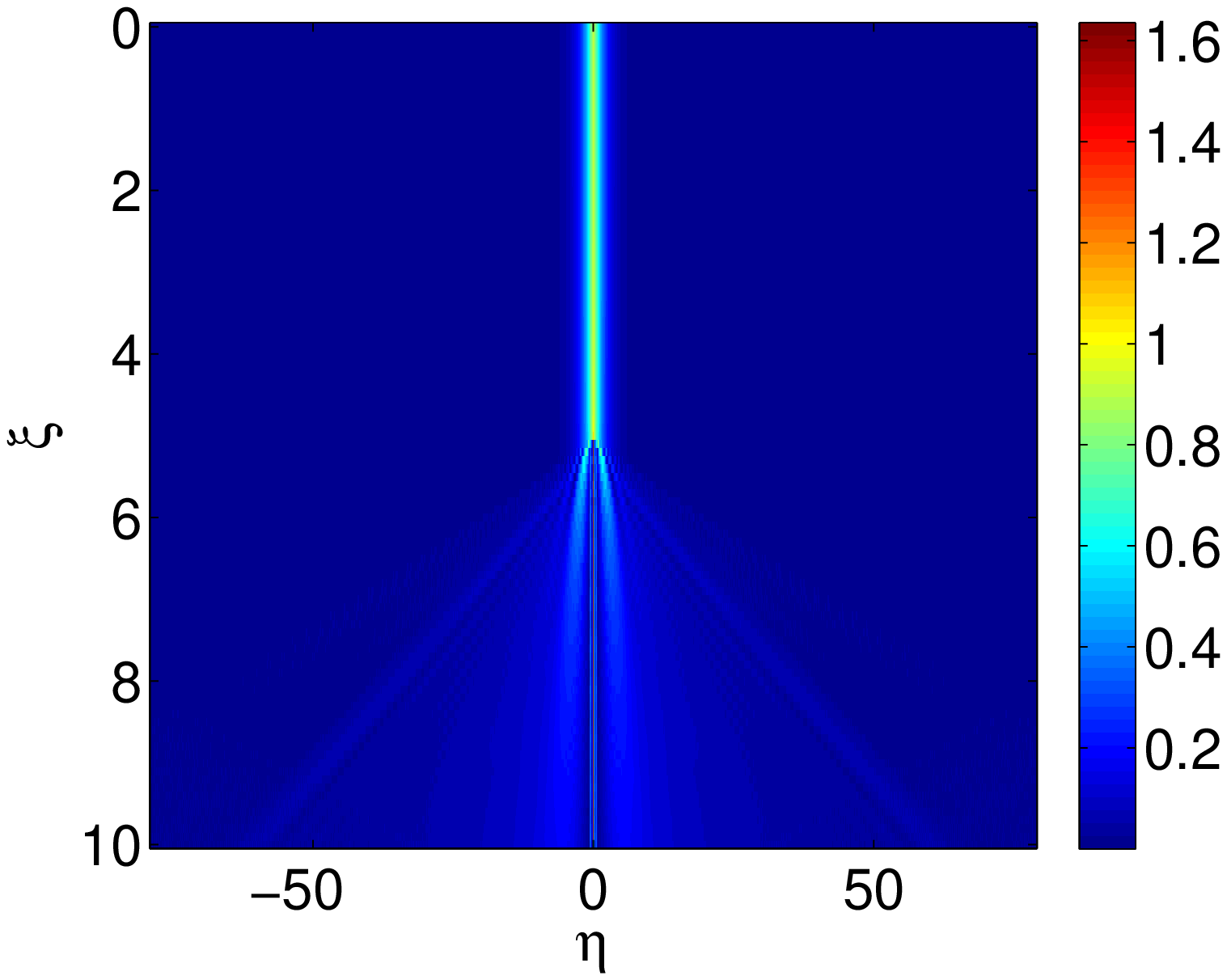} \includegraphics[width=.45%
\textwidth]{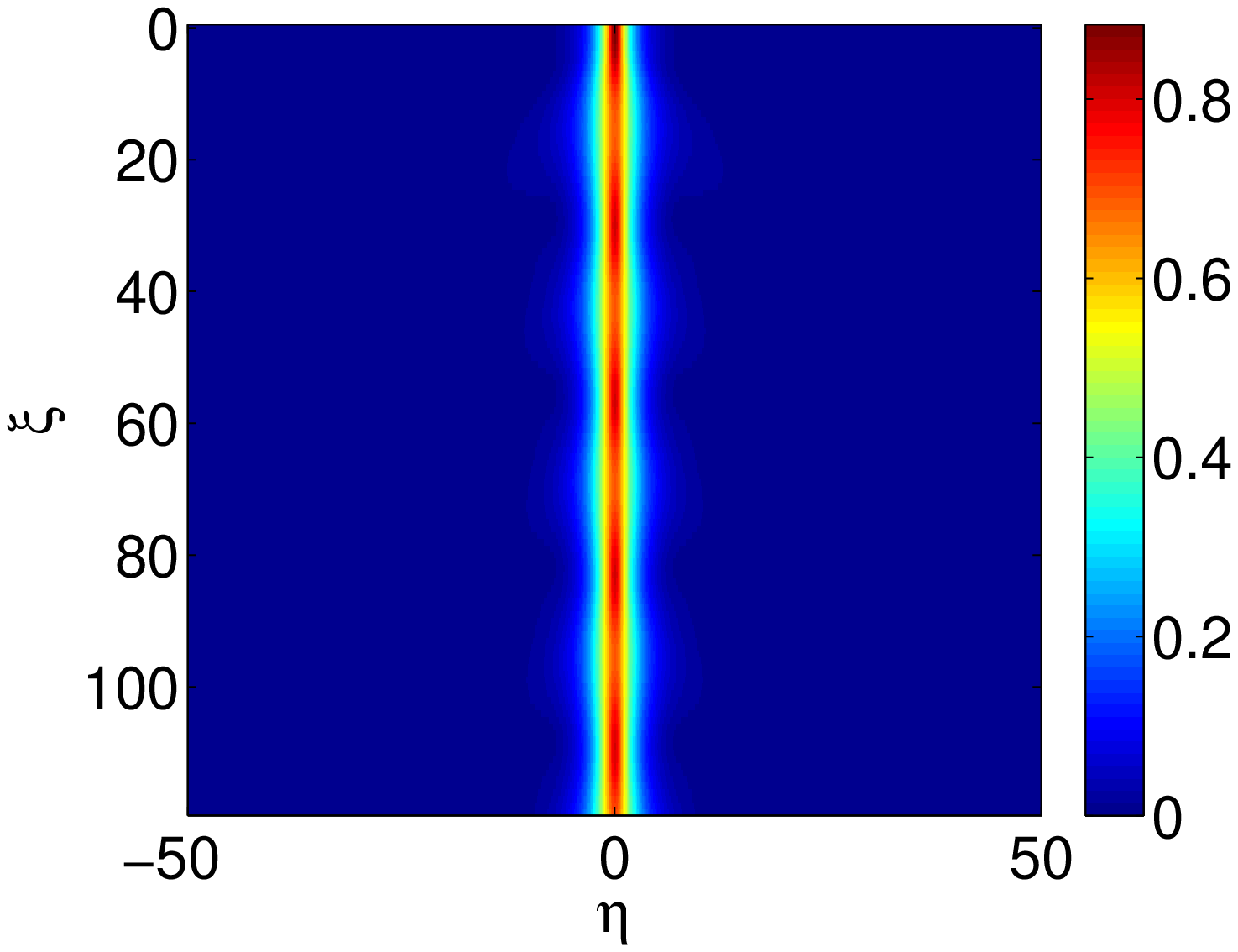} \includegraphics[width=.45%
\textwidth]{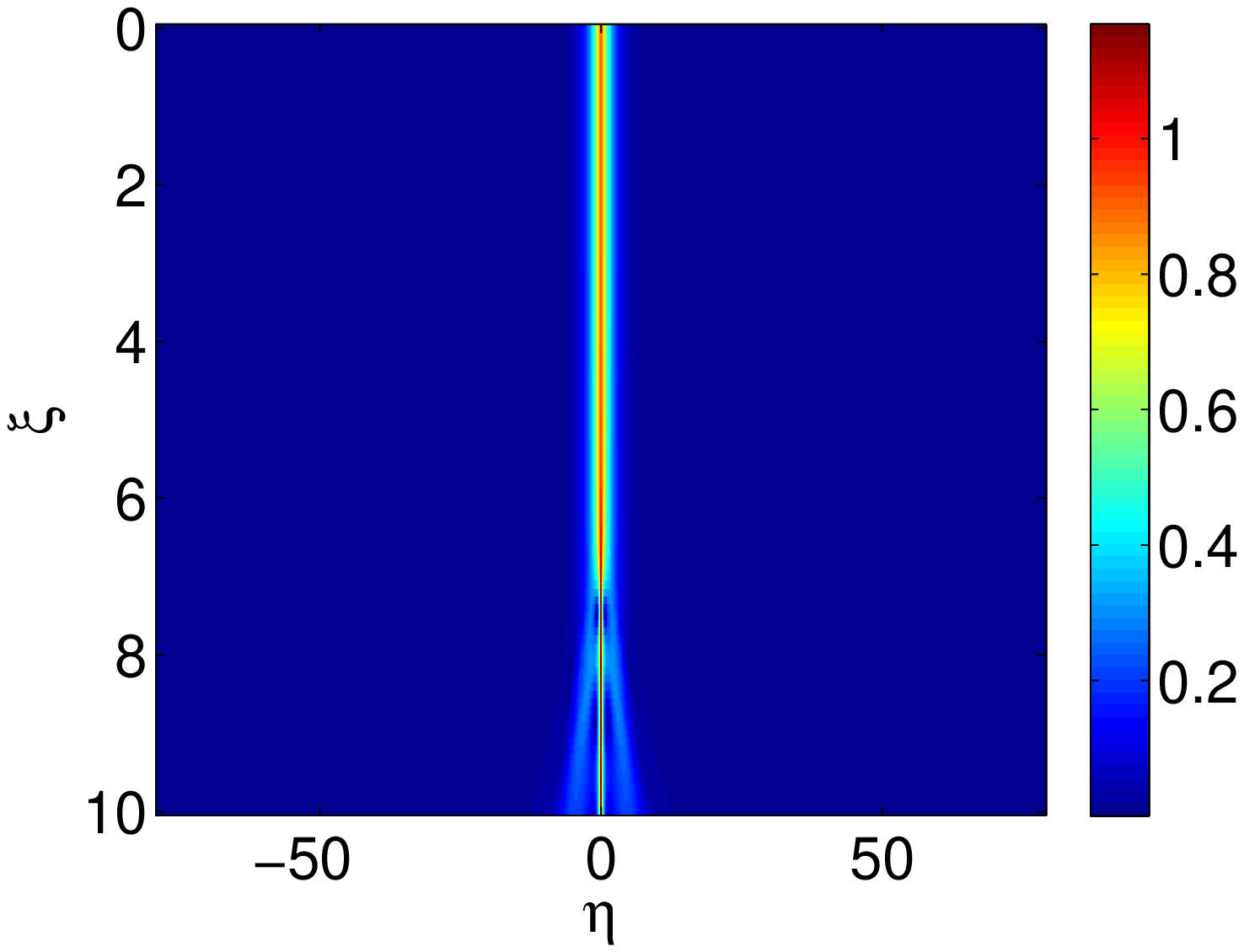} \includegraphics[width=.45%
\textwidth]{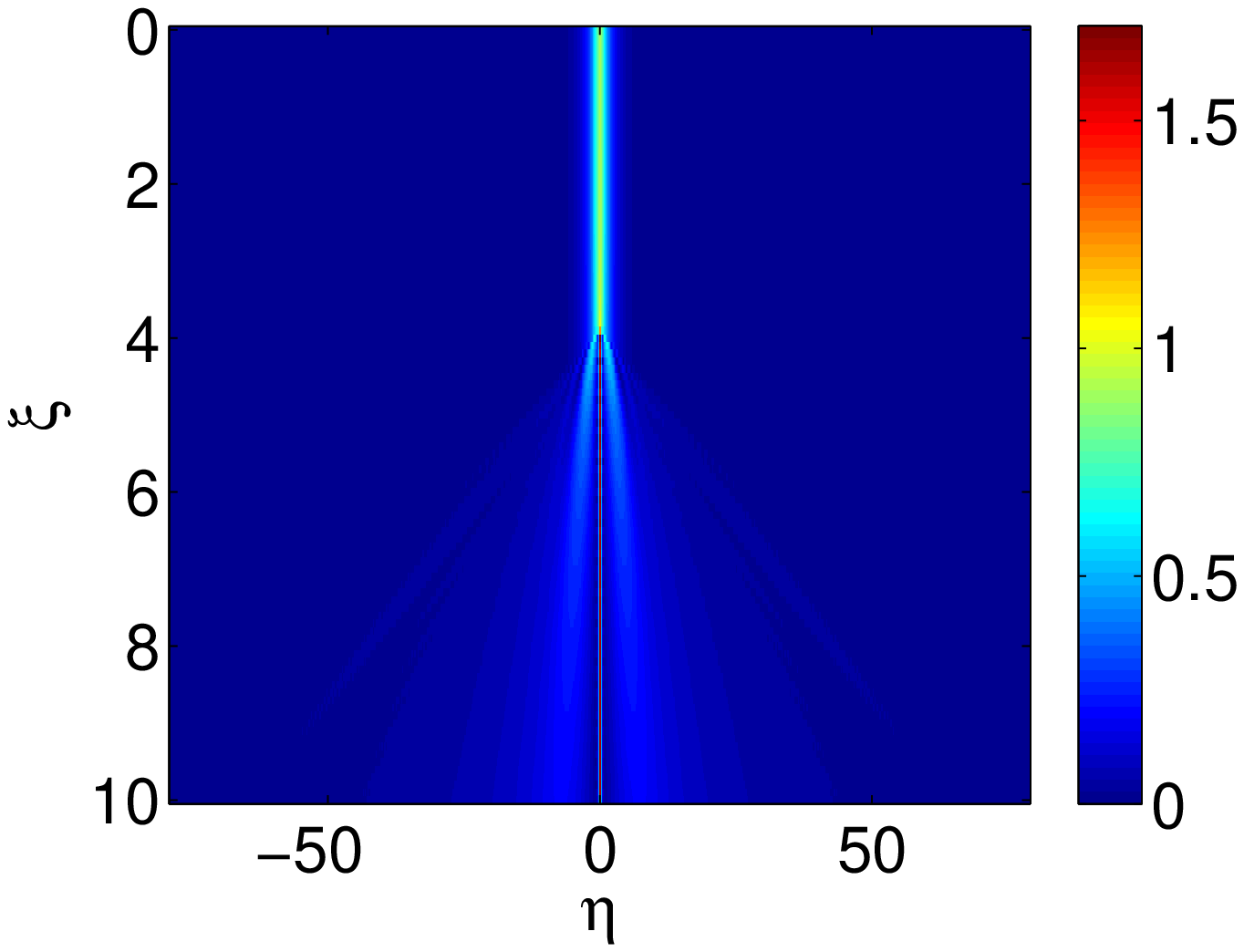}
\caption{(Color online) Contour plots of the evolution of $|q|$ with an
initial perturbation of amplitude $0.001$. The top left panel: a stable
soliton at $k=0.3,\ \protect\beta =1,\ \protect\gamma =1$. The top right
panel: an unstable soliton perturbed by the odd eigenmode at $k=0.4,\
\protect\beta =1,\ \protect\gamma =1$. The middle left panel: the stationary
soliton plus the even perturbation for $k=0.4,\ \protect\beta =1,\ \protect%
\gamma =1$. The middle right panel: the unstable soliton minus the even
perturbation for $k=0.4,\ \protect\beta =1,\ \protect\gamma =1$. The bottom
left panel: the unstable soliton plus the even perturbation for $k=0.4,\
\protect\beta =2,\ \protect\gamma =0$. The bottom right panel: the unstable
soliton plus the even perturbation for $k=0.4,\ \protect\beta =0,\ \protect%
\gamma =2$.}
\label{stabilityFig}
\end{figure}

Predictions of the stability analysis are confirmed by direct numerical
simulations of the evolution of perturbed solitary waves. In the top left
panel of Fig. \ref{stabilityFig}, one observes that, despite adding
perturbations, the solution remains intact for small $k$ (where the solitary
waves are linearly stable). On the contrary, the self-focusing instability,
leading to the above-mentioned \textit{weak collapse}, which involves a
small part of the total power at the center of the beam, and splits the rest
into diverging jets, is evident for a narrower solution with a larger
amplitude (i.e., a larger $k$) in the top right panel of Fig. \ref%
{stabilityFig}. Strictly speaking, the collapse per se is not correctly
described by Eq. (\ref{ele}), as it does not include the fourth-order linear
diffraction (which would ultimately prevent it); however, the equation is
adequate to describe the evolution of the beam \emph{towards }the collapse.

As mentioned before there are two unstable eigenvalues when $k>0.33$. One of
them corresponds to an odd eigenmode, and the other one to an even
eigenmode with a positive amplitude at the center. Numerical simulations
show that, for $k>0.33$, the stationary soliton perturbed by the odd
eigenmode will end up in the weak collapse, see the top right panel in Fig. (%
\ref{stabilityFig}). On the other hand, with the even perturbation, the
dynamics differ for different signs of the perturbations. The stationary
soliton with the even-eigenmode perturbation added to it will collapse, see
the middle left panel in Fig. (\ref{stabilityFig}). On the other hand, under
the action of the perturbation with the opposite sign, the solitary wave
tends to lower its amplitude by emitting small bursts of radiation and
ultimately relaxing into a breather, as is illustrated by the middle right
panel of Fig. (\ref{stabilityFig}). 
 Finally, 
the observation of the effects of the different parameter variations
in the bottom panels of Fig. (\ref%
{stabilityFig}) leads to the conclusion that a higher value of $\gamma $
induces earlier collapse, while the opposite effect is observed with the
increase of $\beta $.

The above simulations started with the input in the form of sech beams. On
the other hand, the VA was based on the Gaussian ansatz (\ref{Gauss}),
therefore it is relevant to test the evolution of solitons generated by the
Gaussian input with parameters predicted by the VA. A typical example, for
parameters belonging to the stability region, is displayed in Fig. \ref%
{Gaussian_input}. One can observe that small (radiation)
jets are shed off at the
initial stage of the evolution---obviously, with the intention to adjust the
Gaussian to the exact solitonic shape--- followed thereafter by the stable
propagation of the resulting wave.
\begin{figure}[tbp]
\centering\includegraphics[width=.45\textwidth]{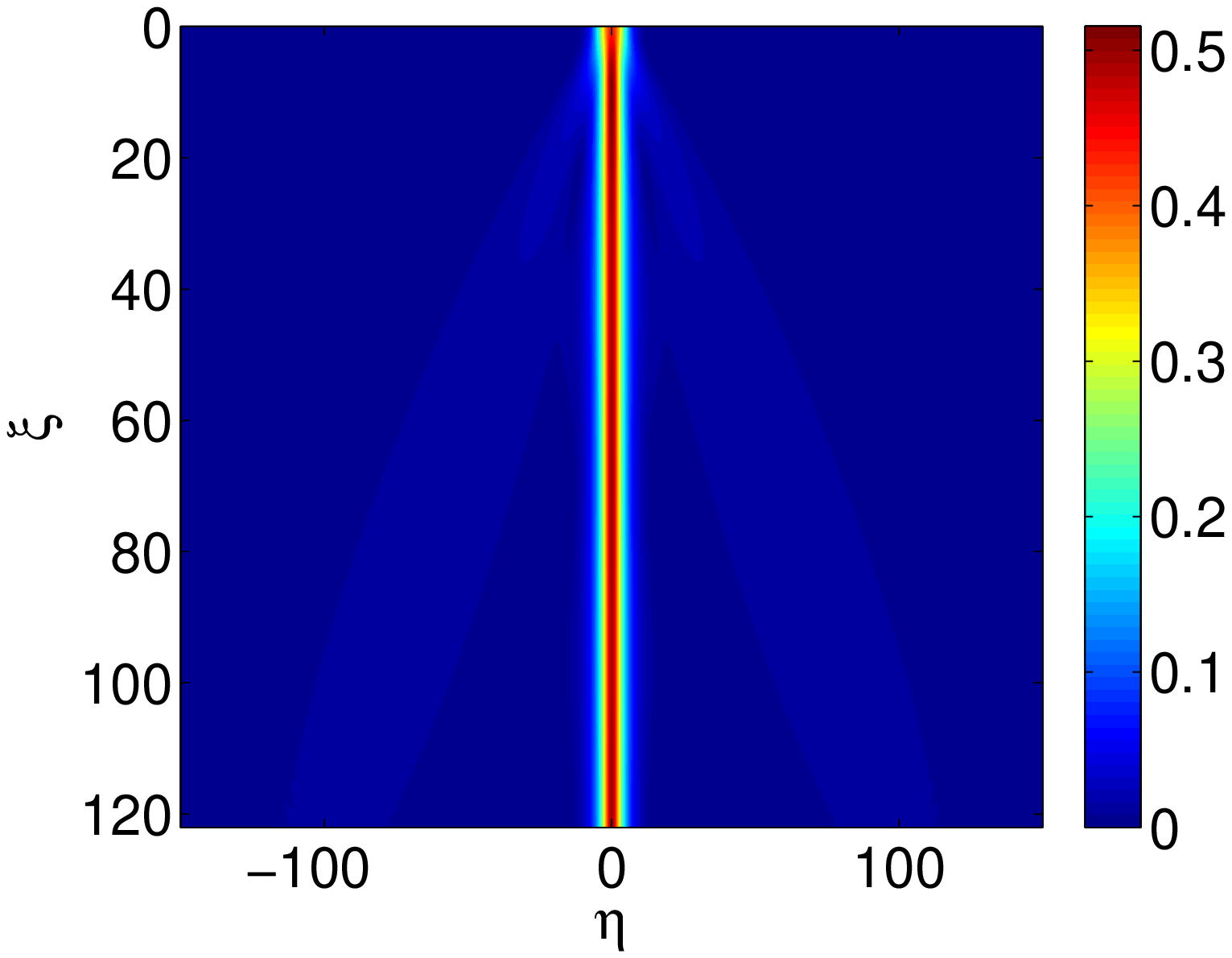}
\caption{(Color online) The same as in Fig. \protect\ref{stabilityFig}, but
for the soliton generated by the Gaussian waveform predicted by the VA, at $%
k=0.1,\ \protect\beta =\protect\gamma =1$.}
\label{Gaussian_input}
\end{figure}

\begin{figure}[tbp]
\par
\begin{center}
\includegraphics[width=.45\textwidth,]{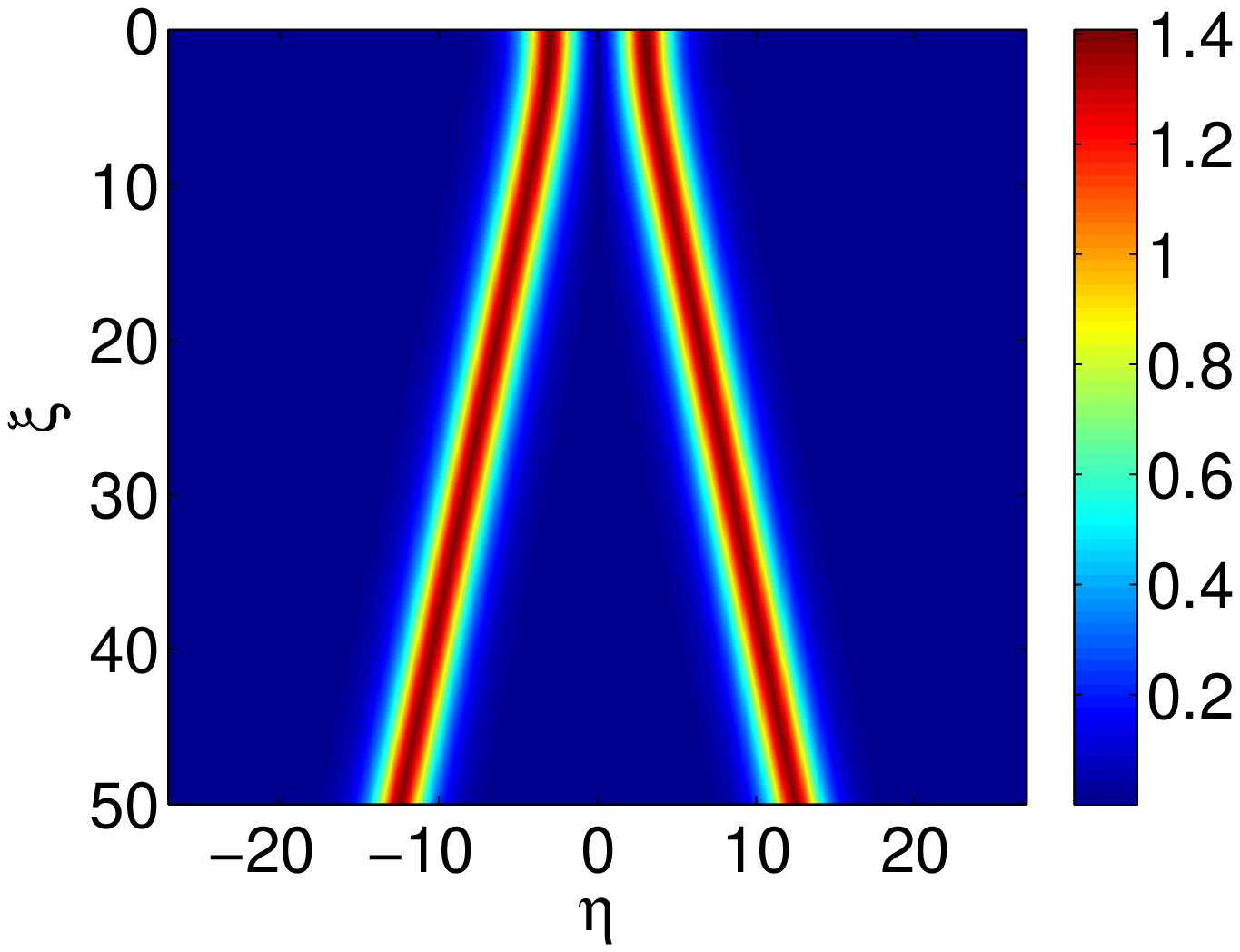}
\end{center}
\par
\centering
\includegraphics[width=.45\textwidth]{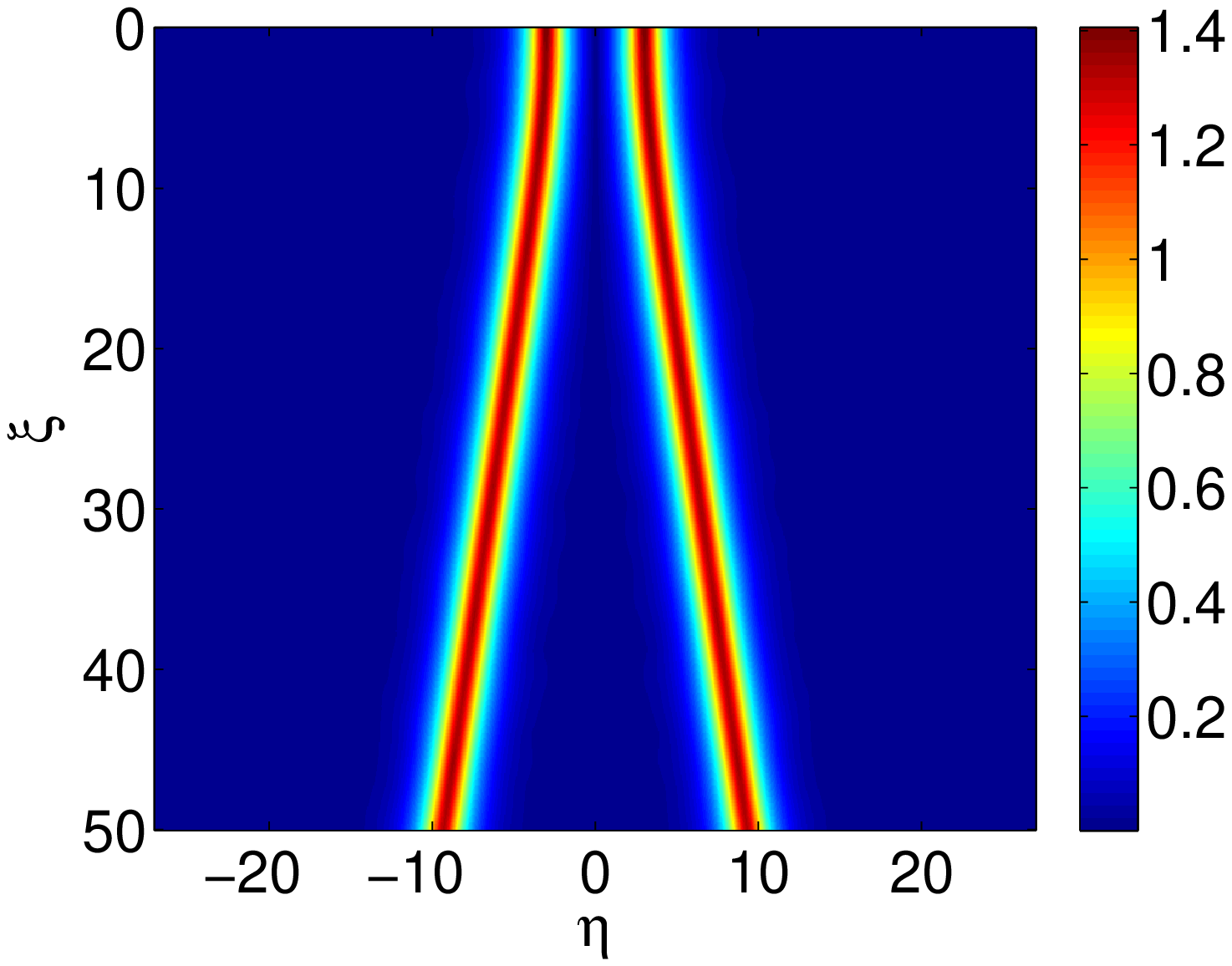} %
\includegraphics[width=.45\textwidth]{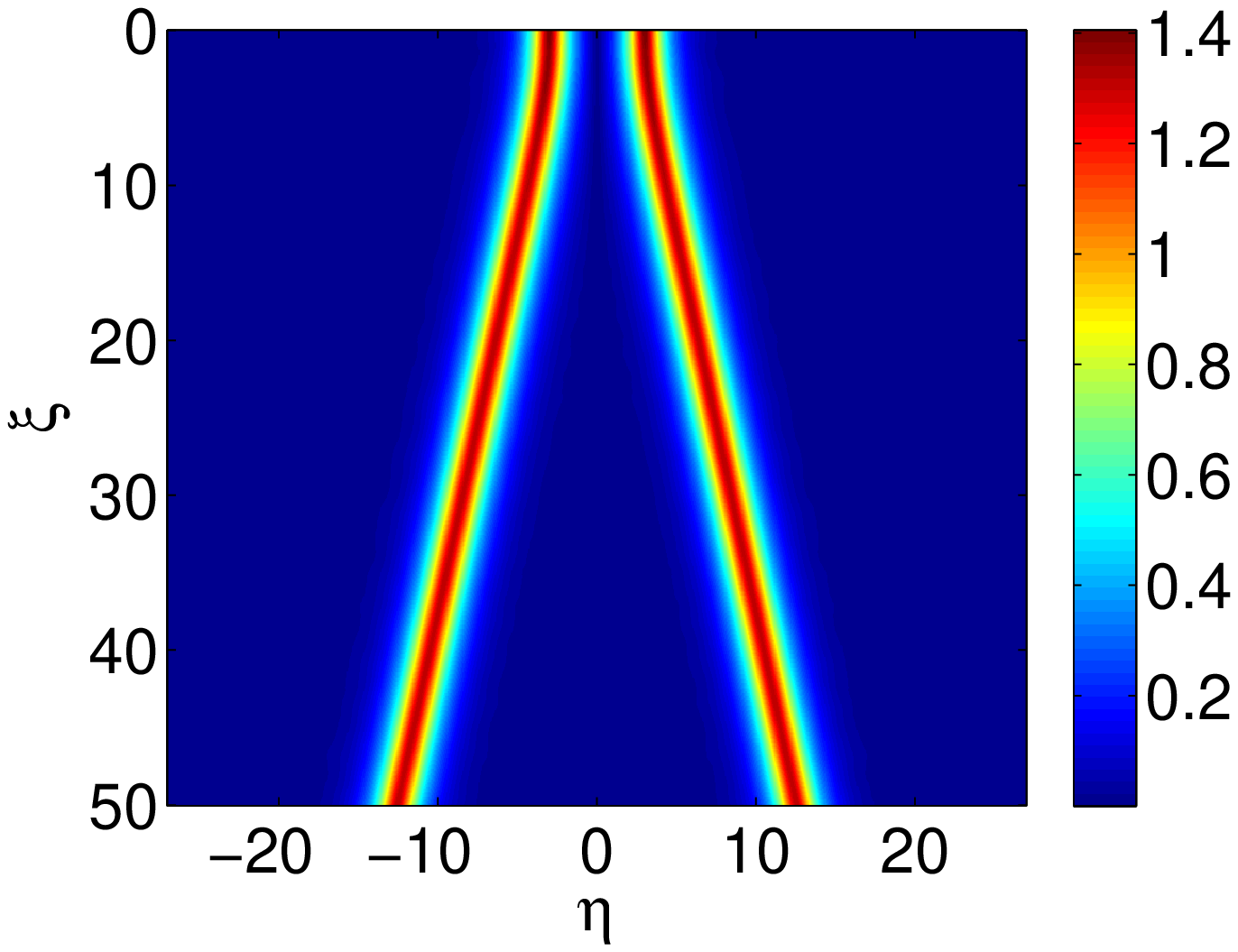}
\caption{(Color online) The repulsive interaction between out-of-phase
solitons. The top panel is for $\protect\beta =0,\protect\gamma =0$, the
bottom left for $\protect\beta =0.6,\protect\gamma =0$, and the bottom right
for $\protect\beta =0,\protect\gamma =0.6$, respectively.}
\label{interaction}
\end{figure}

\begin{figure}[tbp]
\begin{center}
\includegraphics[width=.45\textwidth]{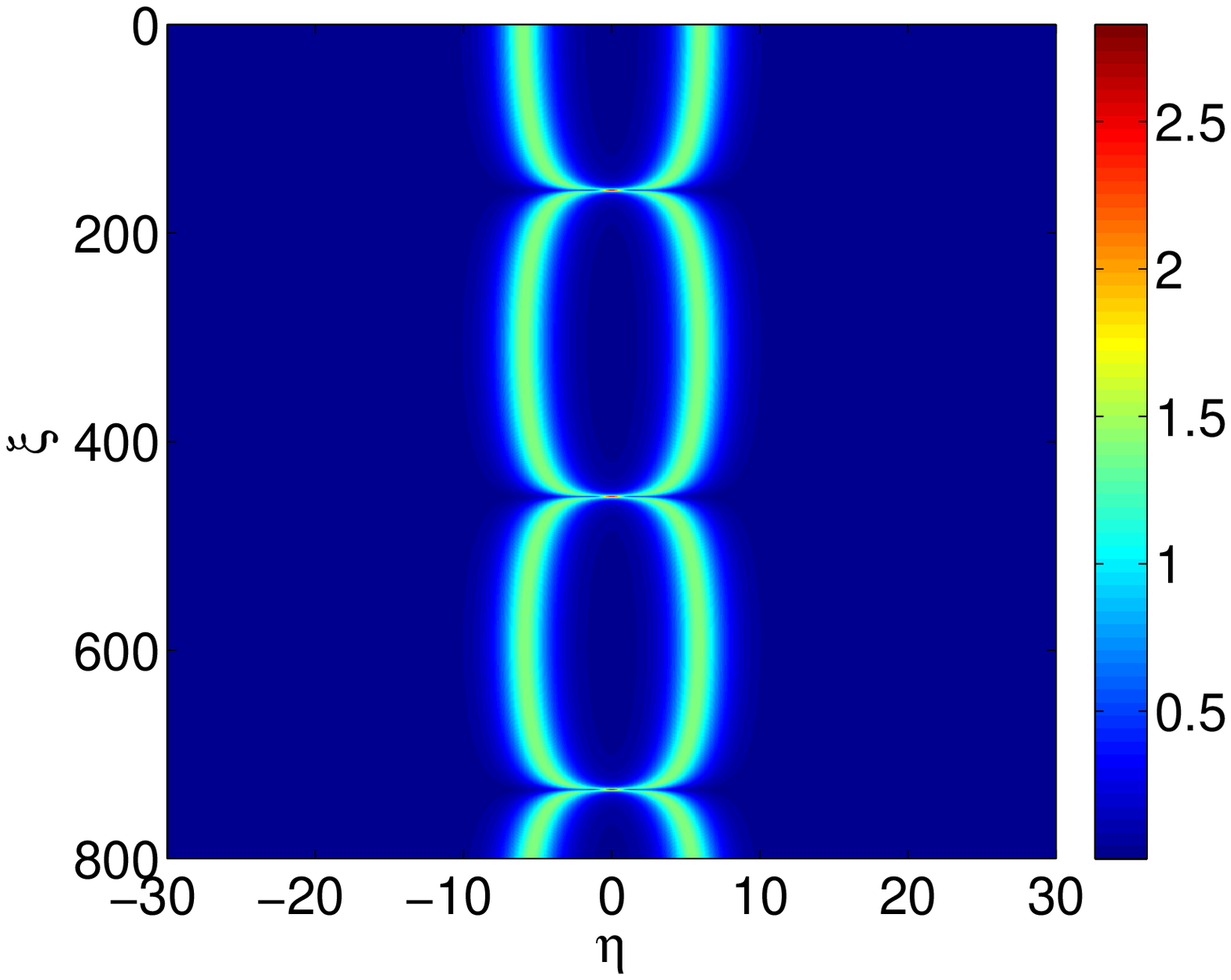}
\end{center}
\par
\centering
\includegraphics[width=.45\textwidth]{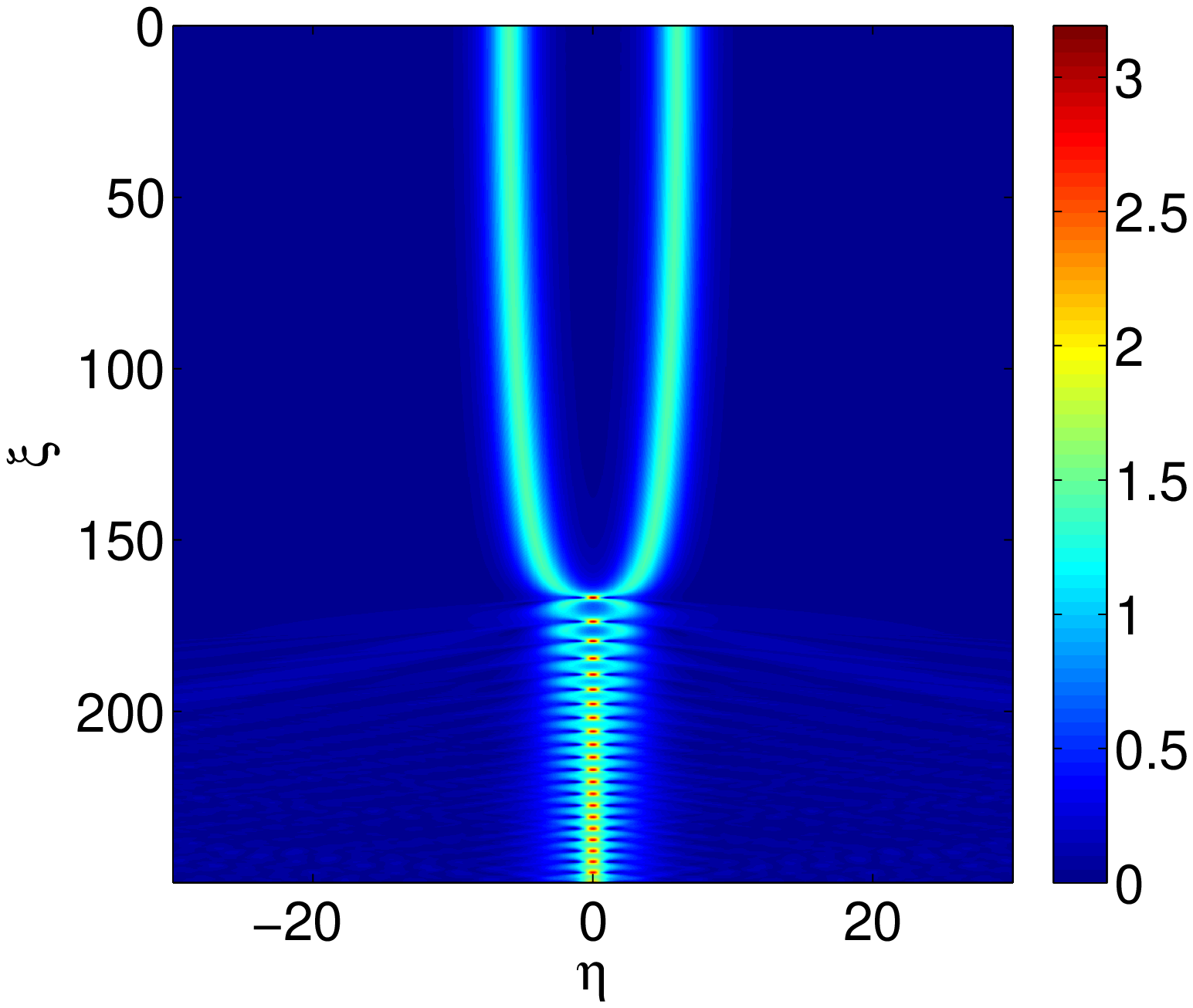}
\includegraphics[width=.45\textwidth]{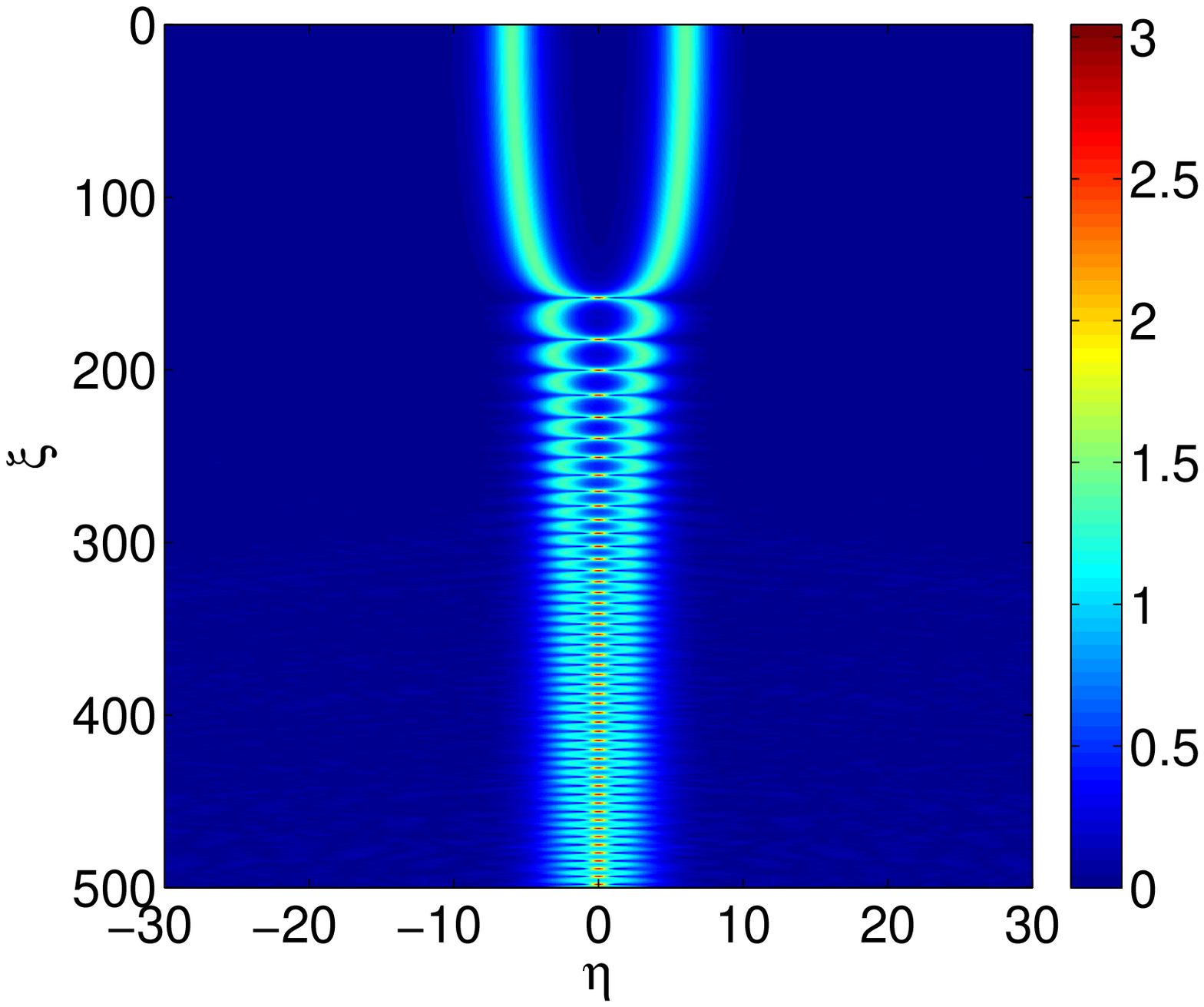}\\
\includegraphics[width=.45\textwidth]{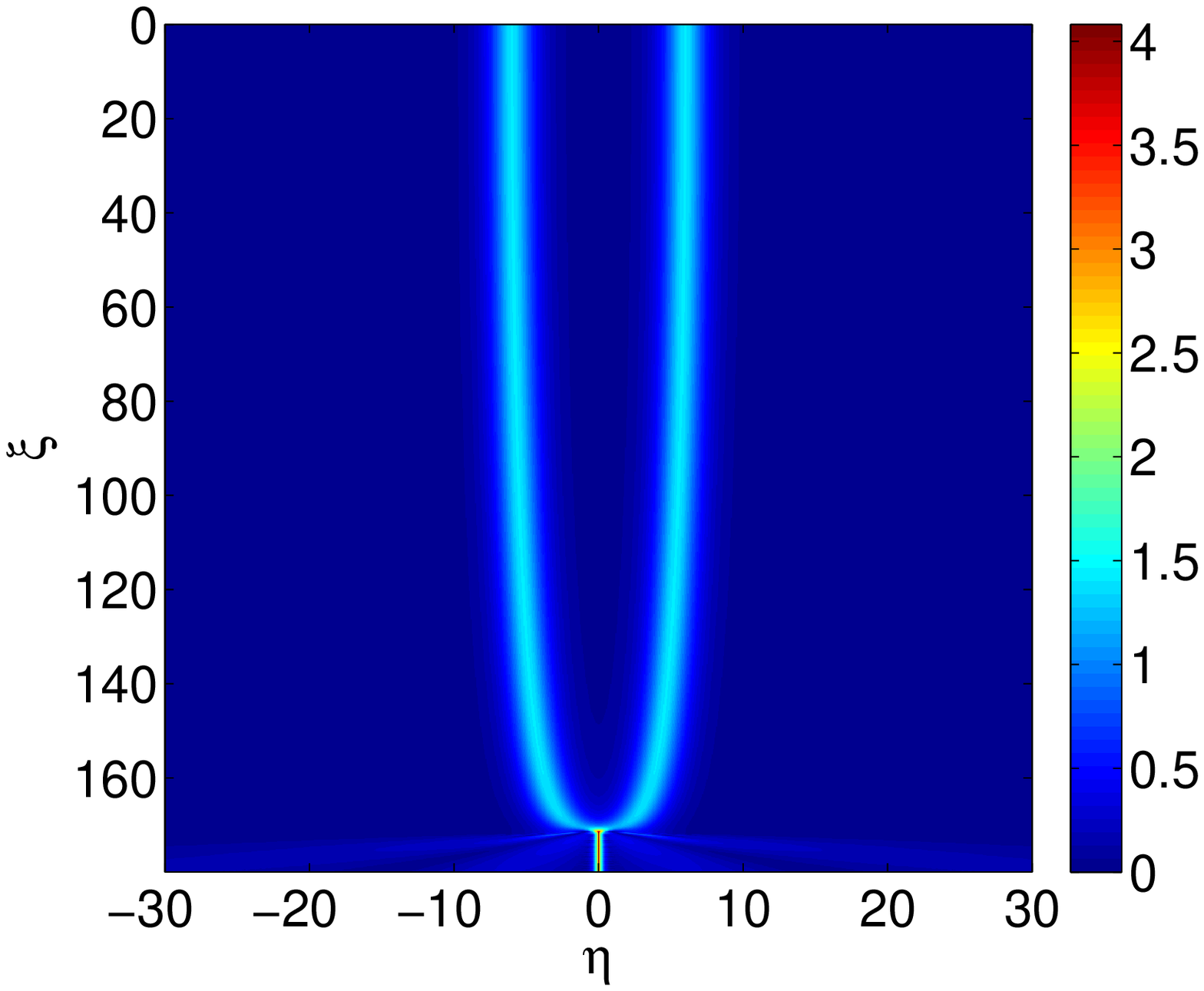}
\includegraphics[width=.45\textwidth]{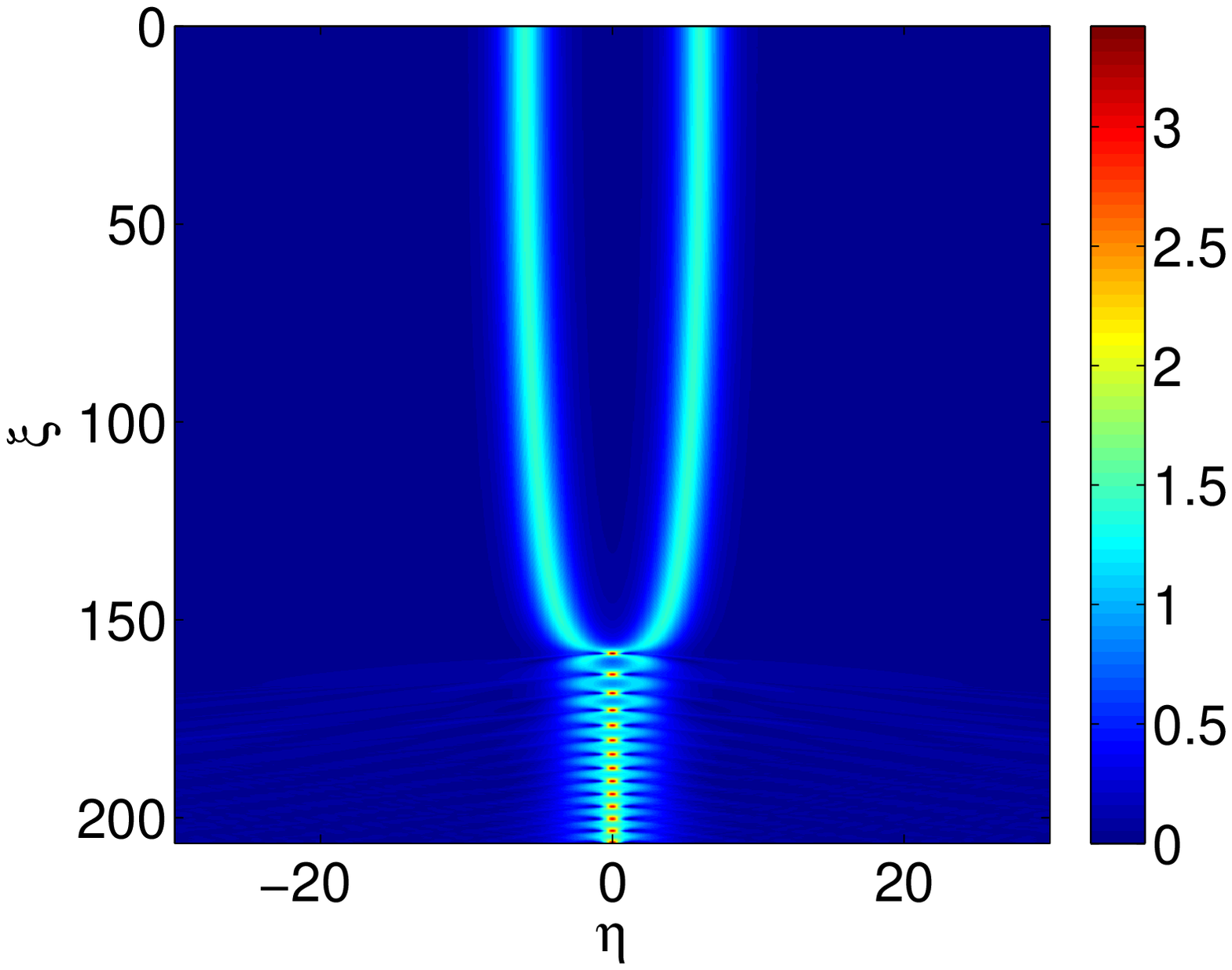}
\caption{(Color online) The panels display the interaction of two
in-phase solitons. The top panel shows periodic elastic collisions
of the soliton pair in the integrable NLS equation, with
$\protect\beta =0,\protect\gamma =0 $ (in fact, this pattern
corresponds to a breather solution of the NLS
equation \protect\cite{breather}). The middle left and right panels are for $%
\protect\beta =0.08,\protect\gamma =0\ $and $\protect\beta =0,\protect\gamma %
=0.08$, respectively. The bottom left and right panels pertain,
respectively,
to $\protect\beta =0.12,\protect\gamma =0$, and $\protect\beta =0,\protect%
\gamma =0.12$.} \label{interaction_critical}
\end{figure}

Finally, we consider the interaction of two solitary beams, for the basic
cases of in-phase and $\pi $-out-of-phase soliton pairs (i.e., mutually
attracting and repelling ones) . The initial in-phase and out-of-phase pairs
without a relative tilt (i.e., the solitons with zero relative velocity)
were built using two well separated copies of a stationary soliton produced
by the Newton's method for $k=1$, while parameters $\beta $ and $\gamma $,
which account for the nonlinear-diffraction terms in Eq. (\ref{ele}), were
varied within boundaries of the stability regime. We observe that, in the
presence of these terms, the repulsive interaction in the out-of-phase
configuration does not change dramatically. However, the two parameters
affect the interaction in different ways. In particular, the increase of $%
\beta $ leads to a decrease in the strength of the repulsion (or attraction)
between the solitons, while the increase of $\gamma $ produces the opposite
effect, see Figs. \ref{interaction} and \ref{interaction_critical}.

While the repulsion, leading to the separation of the solitons in Fig. \ref%
{interaction}, does not reveal any surprising findings, the attractive case
is far more interesting. The nonlinear-diffraction terms lead to the loss of
elasticity of the collisions between NLS solitons, and their eventual merger
into a single pulsating beam. Furthermore, it is possible to identify two
different outcomes of the interaction in this case: The merger may lead to a
\textit{subcritical} scenario, i.e., the appearance of a non-collapsing
robust pulson, that may slowly relax into a stable stationary soliton
through weak radiation losses,\textbf{\ }or a \textit{supercritical} one,
which ends up collapsing. These two cases are represented, respectively, by
the middle and bottom panels in Fig.~\ref{interaction_critical}. In the
subcritical case, it is observed that the merger into the stable pulson is
delayed by the terms $\propto \beta $, or accelerated by the ones $\propto
\gamma $, but the merger happens, eventually. On the other hand, in the case
shown in the bottom row of Fig. \ref{interaction_critical}, the
supercritical merger, leading to the collapse, occurs at $\beta =0.12,\gamma
=0$, while no collapse happens at $\beta =0,\gamma =0.12$
In the latter case, more radiation is observed to be emitted in the course
of the merger, which, in turn, reduces the total power of the emerging
pulson, eventually rendering it subcritical. This is a remarkable feature,
which is uncommon in models that we are aware of.

\section{Conclusion} The objective of
this work was to introduce a generalized model which accounts for phenomena
of nonlinear diffraction within the one-dimensional NLS equation from the
perspective of the self-consistent Lagrangian/Hamiltonian formulation. The
so obtained model contains two independent parameters, $\beta $ and $\gamma $
in Eq. (\ref{ele}). The immediate motivation for the development of the
model is the application to nonlinear photonic crystals in a vicinity of the
supercollimation point, as suggested by Ref. \cite{xu}. In the framework of
the proposed general equation, we have developed a systematic analysis of
the fundamental solitary waves, using the VA (variational approximation),
exact analytical results (wherever possible), and numerical methods, which
corroborate our analytical observations for individual solitons and produce
new results for their pairwise interactions. In particular, it has been
demonstrated that the VK stability criterion is valid for the solitons in
the present model, and the solitons exist only up to a maximum value of the
power. The solitary-wave family consists of stable and unstable parts, the
border between which, $k_{\mathrm{co}}$, and the corresponding largest value
of the total power of the solitons, $P_{\max }$, along with the largest peak
power and smallest width that the stable solitons attain, were found in the
exact form, as written in Eq. (\ref{co}). Interactions between pairs of
identical solitons were studied by means of direct simulations. It has been
demonstrated that the merger of in-phase solitons may lead to the formation
of a robust pulson, if the nonlinear diffraction is not too strong;
otherwise, the emerging beam collapses. On the other hand, the out-of-phase
waves always repel each other and separate, with the interaction timescales
altered by the nonlinear diffraction effects. 


We expect that such a generalized model may hold promise for future
explorations, especially if tuned to experimental data for relevant
photonic crystal structures, so that the relative strength of 
the parameters considered herein can be evaluated in realistic
settings. On the other hand, it would be particularly relevant
to extend the present considerations to other regimes including
the ones pertaining to self-defocusing nonlinearities, or to
higher-dimensional photonic crystals (possibly with either
sign of the nonlinearity). Such studies will be deferred to
future publications.

\acknowledgments PGK gratefully acknowledges M. Solja\v{c}i\'{c}
for originally bringing this problem to his attention, as
well as Ref.~\cite{xu} and for helpful discussions on this theme.
PGK also gratefully acknowledges support from the National
Science Foundation under grant DMS-0806762. PGK and BAM also
acknowledge support from the Binational Science Foundation
under grant 2010239.

\begin{thebibliography}{99} 
\bibitem{management} H. S.
Eisenberg, Y. Silberberg, R. Morandotti, and J. S. Aitchison, Phys. Rev.
Lett. \textbf{85}, 1863 (2000); M. J. Ablowitz and Z. H. Musslimani, \textit{%
ibid}. \textbf{87}, 254102 (2001); A. A. Sukhorukov and Y. S. Kivshar, Opt.
Lett. \textbf{27}, 2112 (2002); N. K. Efremidis and D. N. Christodoulides,
Phys. Rev. E \textbf{65}, 056607 (2002); A. A. Sukhorukov, Y. S. Kivshar, H.
S. Eisenberg, and Y. Silberberg, IEEE J. Quant. Electr. \textbf{39}, 31
(2003); P. G. Kevrekidis, B. A. Malomed, and Z. Musslimani, Eur. Phys. J. D
\textbf{23}, 421 (2003); P. G. Kevrekidis, B. A. Malomed, A. Saxena, A. R.
Bishop, and D. J. Frantzeskakis, Physica D 183, 87 (2003); N. K. Efremidis
and K. Hizanidis, Opt. Exp. \textbf{13}, 10571 (2005); Y. V. Kartashov, L.
Torner, and V. A. Vysloukh, Opt. Exp. \textbf{13}, 4244 (2005); J. T.
Moeser, Nonlinearity \textbf{18}, 2275 (2005); P. Panayotaros, Phys. Lett.
\textbf{349}, 430 (2006); A. D. Boardman, R. C. Mitchell-Thomas, N. J. King,
and Y. G. Rapoport, Opt. Commun. \textbf{283}, 1585 (2010).

\bibitem{Zhong}
W.-P. Zhong, M. Beli\'{c}, G. Assanto, B. A Malomed, and T. Huang, Phys.
Rev. A \textbf{84}, 043801 (2011). 

\bibitem{Barcelona} O. V. Borovkova, Y.
V. Kartashov, B. A. Malomed, and L. Torner, Opt. Lett. \textbf{36}, 3088
(2011); O. V. Borovkova, Y. V. Kartashov, L. Torner, and B. A. Malomed,
Phys. Rev. E 84, 035602(R) (2011); Y. V. Kartashov, V. A. Vysloukh, L.
Torner, and B. A. Malomed, Opt. Lett. \textbf{36}, 4587 (2011). 

\bibitem{%
experiment} H. Kosaka, T. Kawashima, A. Tomita, M. Notomi, T. Tamamura, T.
Sato, and S. Kawakami, Appl. Phys. Lett. \textbf{74}, 1212 (1999); D. M.
Pustai, S. Shi, C. Chen, A. Sharkawy, and D. W. Prather, Opt. Express
\textbf{12}, 1823 (2004); P. T. Rakich, M. S. Dahlem, S. Tandon, M.
Ibanescu, M. Solja\v{c}i\'{c}, G. S. Petrich, J. D. Joannopoulos, L. A.
Kolodziejski, and E. P. Ippen, Nature Materials \textbf{5}, 93 (2006);

\bibitem{theory} M. Notomi, Phys. Rev. B \textbf{62}, 10696 (2000); J.
Witzens and A. Scherer, J. Opt. Soc. Am. A \textbf{20}, 935 (2003); J. Shin
and S. Fan, Opt. Lett. \textbf{30}, 2397 (2005); X. Y. Jiang, C. H. Zhou, X.
F. Yu, S. H. Fan, M. Solja\v{c}i\'{c}, and J. D. Joannopoulos, Appl. Phys.
Lett. \textbf{91}, 031105 (2007). 

\bibitem{Kestas} K. Staliunas and R.
Herrero, Phys. Rev. E \textbf{73}, 016601 (2006); R. Iliew, C. Etrich, T.
Pertsch, E. Lederer, and K. Staliunas, Opt. Lett. \textbf{33}, 2695 (2008).

\bibitem{xu} Z. Xu, B. Maes, X. Jiang, J. D. Joannopoulos, L. Torner and M.
Solja\v{c}i\'{c}, Opt. Lett. \textbf{33}, 1762 (2008). 

\bibitem{arrays} K.
Staliunas and C. Masoller, Opt. Exp. 14, 10669 (2006); S. Longhi and K.
Staliunas, Opt. Commun. \textbf{281}, 4343 (2008). \bibitem{cavity} O.
Egorov, F. Lederer, and K. Staliunas, Opt. Lett. \textbf{32}, 2106 (2007).

\bibitem{quadr-array} O. Egorov and F. Lederer, Opt. Exp. \textbf{16}, 6050
(2008). 

\bibitem{sonic-crystal} V. Espinosa, V. J. Sanchez-Morcillo, K.
Staliunas, I. Perez-Arjona, and J. Redondo, Phys. Rev. B \textbf{76}, 140302
(2007). \bibitem{BEC} K. Staliunas, R. Herrero, and G. J. de Valcarcel,
Phys. Rev. E \textbf{73}, 065603 (2006). 

\bibitem{VA} B. A. Malomed, Progr.
Optics \textbf{43}, 71 (E. Wolf, editor: North Holland, Amsterdam, 2002);
http://www.sciencedirect.com/science/article/pii/S0079663802800269. 

\bibitem{%
Zaragoza} J. Gomez-Garde\~{n}es, B. A. Malomed, L. M. Floria, and A. R.
Bishop, Phys. Rev. E \textbf{73}, 036608 (2006). \bibitem{additional} M.
Marklund, P. K. Shukla, R. Bingham, and J. T. Mendon\c{c}a, Phys. Rev. A
\textbf{74}, 045801 (2006). 

\bibitem{Andrei} A. I. Maimistov and Yu. M.
Sklyarov, Kvant. Elektronika \textbf{14}, 796 (1097) [in Russian; English
translation: Sov. J. Quant. Electron. \textbf{17}, 500 (1987). \bibitem{VK}
M. Vakhitov and A. Kolokolov, Radiophys. Quantum. Electron. \textbf{16}, 783
(1973). 

\bibitem{Berge} L. Berg\'{e}, Phys. Rep. \textbf{303}, 259 (1998).
\bibitem{Kuz} E. A. Kuznetsov, Chaos \textbf{6}, 381 (1996); E. A. Kuznetsov
and F. Dias, Phys. Rep. \textbf{507}, 43 (2011). 

\bibitem{breather} Yu. S.
Kivshar and B. A. Malomed, Rev. Mod. Phys. \textbf{61}, 763 (1989). \end{%
thebibliography} \end{document}